\documentclass[12pt,modern]{aastex631}
\date{Astrophysical Journal, Re-submitted March 13, 2025}

\usepackage{verbatim}
\usepackage{natbib,aas_macros}
\usepackage{hyperref}
\usepackage{graphicx}
\usepackage{xspace}
\usepackage{xcolor}  
\usepackage{rotating} 
\usepackage{comment}

\newcommand{\cchp}{\,CCHP\xspace}                        
\newcommand{\csp}{\,CSP\xspace}   
\newcommand{\cspi}{\,CSP-I\xspace}                      
\newcommand{\cspii}{\,CSP-II\xspace} 

\newcommand{\gaia}{\emph{Gaia}\xspace}            
\newcommand{\ho}{$H_{0}$\xspace}                  
\newcommand{\hounits}{\,km\,s$^{-1}$\,Mpc$^{-1}$\xspace}   
\newcommand{\hst}{\emph{HST}\xspace}              
\newcommand{\hstwfciii}{\emph{HST/WFC3}\xspace}    
\newcommand{\hstacs}{\emph{HST/ACS}\xspace} 
\newcommand{\jagb}{{\it JAGB}\xspace} 
\newcommand{\jwst}{\emph{JWST}\xspace}

\newcommand{\nircam}{\emph{NIRCam}\xspace}  
  
\newcommand{\planck}{\emph{Planck}\xspace}
            
\newcommand{\sne}{SNe~Ia\xspace}                   
\newcommand{\sn}{SN~Ia\xspace}                   
               
\newcommand{\shoes}{\emph{SHoES}\xspace}            
\newcommand{\ngc}{NGC\,\xspace}
\newcommand{\ic}{IC\,1613\xspace}

\begin{document}

\shorttitle{JWST Hubble Program}
\shortauthors{Freedman et al.}

\title{Status Report on the Chicago-Carnegie Hubble Program (CCHP): \\ Measurement of the Hubble Constant Using \\ the Hubble and James Webb Space Telescopes 
\protect\footnote{This work is based on observations made with the NASA/ESA/CSA James Webb Space Telescope. The data were obtained from the Mikulski Archive for Space Telescopes at the Space Telescope Science Institute, which is operated by the Association of Universities
for Research in Astronomy, Inc., under NASA contract NAS 5-03127 for JWST. These observations are associated with program JWST-GO-1995.}}

\bigskip

\correspondingauthor{Wendy Freedman}
\email{wfreedman@uchicago.edu}

\author[0000-0003-3431-9135]{Wendy~L.~Freedman}
\affiliation{Department of Astronomy \& Astrophysics, University of Chicago, 5640 South Ellis Avenue, Chicago, IL 60637, USA}\affiliation{Kavli Institute for Cosmological Physics, University of Chicago,  5640 South Ellis Avenue, Chicago, IL 60637}

\author[0000-0002-1576-1676]{Barry~F.~Madore}
\affil{The Observatories of the Carnegie Institution for Science, 813 Santa Barbara St., Pasadena, CA 91101, USA}

\author[0000-0001-9664-0560]{Taylor J. Hoyt}
\affiliation{Physics Division, Lawrence Berkeley National Lab, 1 Cyclotron Road, Berkeley, CA 94720, USA}

\author{In Sung Jang}\affil{Department of Astronomy \& Astrophysics, University of Chicago, 5640 South Ellis Avenue, Chicago, IL 60637, USA}\affiliation{Kavli Institute for Cosmological Physics, University of Chicago,  5640 South Ellis Avenue, Chicago, IL 60637}

\author[0000-0002-5865-0220]{Abigail~J.~Lee}\altaffiliation{LSSTC DSFP Fellow}\affiliation{Department of Astronomy \& Astrophysics, University of Chicago, 5640 South Ellis Avenue, Chicago, IL 60637, USA}\affiliation{Kavli Institute for Cosmological Physics, University of Chicago,  5640 South Ellis Avenue, Chicago, IL 60637}

\author[0000-0003-3339-8820]{Kayla~A.~Owens}\affil{Department of Astronomy \& Astrophysics, University of Chicago, 5640 South Ellis Avenue, Chicago, IL 60637, USA}\affiliation{Kavli Institute for Cosmological Physics, University of Chicago,  5640 South Ellis Avenue, Chicago, IL 60637}

  
\begin{abstract}

We present the latest results from the Chicago-Carnegie Hubble Program (\cchp) to measure the Hubble constant, using data  from the James Webb Space Telescope (\jwst). The overall program aims to calibrate three independent methods:  (1) Tip of the Red Giant Branch (TRGB) stars, (2) JAGB (J-Region Asymptotic Giant Branch) stars, and (3) Cepheids. To date, our program includes 10 nearby galaxies,  hosting 11 Type Ia supernovae (\sne) suitable for measuring the Hubble constant (\ho). It also includes the galaxy \ngc 4258, whose   geometric distance provides the zero-point calibration. In this paper we discuss our results from the TRGB and JAGB methods. Our current best (highest precision) estimate is  \ho = 70.39 $\pm$ 1.22 (stat) $\pm$ 1.33 (sys) $\pm$ 0.70 ($\sigma_{SN})$, based on the TRGB method alone, with a total of 24 \sn calibrators from both \hst and \jwst data.  Based on our new \jwst data only, and tying into \sne, we find values of \ho = 68.81 $\pm$  1.79 (stat) $\pm$ 1.32 (sys) for the TRGB, and   \ho = 67.80  $\pm$ 2.17   (stat) $\pm$ 1.64 (sys) \hounits for the JAGB method.    The distances measured using the TRGB and the JAGB method agree, on average, at a level better than 1\%, and with the \shoes Cepheid distances at just over the 1\% level. Our results are consistent with the current standard $\Lambda$CDM model, without the need for the inclusion of additional new physics. Future \jwst data will be required to increase the  precision and accuracy of the local distance scale.

\end{abstract}

\section{Introduction}
\label{sec:intro}

The year 2025 marks  a century since Edwin Hubble's first publication of the discovery of  Cepheid variables in an external galaxy, \ngc 6822 \citep{hubble_1925}.
Hubble's subsequent measurements of extragalactic distances \citep{hubble_1929} were based, in part, on the Cepheid period-luminosity (PL) relation (also now widely known as the Leavitt Law, \cite{leavitt_1907}). For 70 years a number of unrecognized challenges (e.g., reddening and extinction/dimming due to the effects of interstellar dust, errors in photometric zero points, effects due to differing metal abundances, crowding/blending as a result of insufficient resolution, and the inclusion of some (secondary) distance indicators) turned out to have large systematic effects, and combined together made it  virtually impossible for the Hubble constant (\ho) to be measured from the ground to better than a factor of two uncertainty. 

This impasse was largely overcome by technological advances:
(1) the widespread availability of  two-dimensional, linear detectors operating in the optical and near-infrared, beginning in the 1980s \citep[e.g.][]{mcgonegal_1982, freedman_grieve_madore_1985, freedman_wilson_madore_1991}, and (2) ultimately the launch of the Hubble Space Telescope (\hst), and undertaking of the  Hubble Key Project \citep{freedman_2001}. These order-of-magnitude advances along three axes of sensitivity, wavelength coverage and angular resolution, made it possible to reduce the 100\% (`factor-of-two') uncertainty on the Hubble constant down to 10\%, yielding \ho = 72 $\pm$ 3 (stat) $\pm$ 7 [sys] (\citeauthor{freedman_2001}). 
Over the following decades this space-based measurement of \ho has been confirmed by multiple subsequent analyses \citep[e.g.,][]{riess_2009, freedman_2012, riess_2016, riess_2022}, all based on a Cepheid calibration of distant \sne, subsequently taken into the more distant Hubble flow.

Twenty-five years later, one of the outstanding questions in cosmology is this: Have we overcome the systematic effects (both known and perhaps still unknown) that can affect measurements of the astrophysical distance scale at a sufficient level to require that our current standard cosmological model ($\Lambda$CDM) might now be in need of additional physics? This question has arisen with the emergence of a completely new method for inferring the value of \ho that is totally independent of the Cepheid calibration, and is, instead, based upon modeling measurements of the fluctuations in the cosmic microwave background. Under the assumption of the standard ($\Lambda$CDM) cosmological model, the CMB measurements from the \planck satellite predict a current expansion rate of 67.4 $\pm$ 0.5 \hounits (i.e., with better than 1\% precision)\citep{planck_2018}. Consistent results are also obtained from the Atacama Cosmology Telescope (ACT, \cite[e.g.,][]{ACT_2024}), and from the South Pole Telescope (SPT, \cite[e.g.,][]{Balkenhol_2023, ge_2024}).  
However, the CMB results differ from some recent (local) measurements of \ho based on distant Type Ia supernovae (\sne) in the unperturbed Hubble flow, calibrated using \hst observations of Cepheids. The difference is at a level of 5$\sigma$, as estimated by \cite[][hereafter R22]{riess_2022}, a discrepancy known as the Hubble tension. However,  \hst TRGB measurements \citep{freedman_2019} show no significant tension with the CMB. If a large discrepancy is confirmed, it suggests the existence of new physics (particles or fields) not yet constrained by the  standard ($\Lambda$CDM) cosmological model. To date, however, no plausible changes or additions have emerged that allow for values of \ho  as high as 73 \hounits  \citep[see][]{divalentino_2021c}.

Other physics-based methods, consistent with CMB measurements and lower values of \ho, are also attained from measurements of baryon acoustic oscillations (BAO), or fluctuations in the matter density.  The Dark Energy Spectroscopic Instrument (DESI) Collaboration Data Release 1 (DR1) \citep{DESI_2024} find \ho = 68.52 $\pm$ 0.62 \hounits. This value was obtained by calibrating the standard BAO ruler based on a prior for the baryon density from Big Bang Nucleosynthesis (BBN), in addition to a  measurement of the CMB acoustic angular scale, $\Theta_*$. (The CMB acoustic angular scale has been measured to a very high precision of 0.03\% \citep{planck_2018}). A slightly less precise (but nonetheless the same) result is found using the BBN calibration alone, giving \ho = 68.53 $\pm$ 0.80 \hounits.  While not completely independent of the CMB measurements (these results assume the $\Lambda$CDM model), they offer an important consistency check. If instead, a calibration based on CMB measurement of the sound horizon, r$_d$ is used, the resulting value of \ho is 69.29 $\pm$ 0.87 \hounits (again assuming the $\Lambda$CDM model). Most recently, and also based on the DESI data, combined with additional sound-horizon independent data sets,  \citet{zaborowski_2024} find $66.7^{+1.7}_{-1.9}, \ 67.9^{+1.9}_{-2.1}, \ \text{and} \ 67.8^{+2.0}_{-2.2}$ \hounits. 

Although modeling of gravitationally lensed supernovae is in its early stages,  measurements of the  well-modeled) Supernova Refsdal have yielded low values of \ho = 64.8 $\pm$ 4.4 and 66.6 $\pm$ 4.4 \hounits \citep{kelly_2023}; \ho = 65.1$_{-3.4}^{+3.5}$ \hounits \citep{grillo_2024}\footnote{A second gravitational lens SN H0pe \citep{pascale_2025} currently has almost twice the uncertainty  $\sim$9\%.}. These several, independent (physics-based) methods (CMB, BAO, lensed supernovae) provide no evidence for new early-Universe physics.

Beyond the challenges of calibrating an absolute distance scale, accurate local \ho measurements  based on \sne necessitate precise calibration of \sn peak absolute magnitudes (M$_B$). 
A well-established degeneracy exists between M$_B$ and H$_0$ \citep[e.g.,][]{efstathiou_2021, dainotti_2025}. In essence, any error in M$_B$ whether stemming from measurements of nearby \sne or the distances to Cepheids or TRGB stars used for calibration, directly propagates into an error in the measured \ho value. Literature values for M$_B$ range from -19.253 $\pm$ 0.027 \citep{riess_2022} calibrated by Cepheids,  to -19.396 $\pm$0.016 \citep{dinda_banerjee_2023}, based on a combination of \sne and BAO observations. A calibration of M$_B$ using the TRGB yields M$_B$ = -19.326 $\pm$ 0.038 mag \citep{freedman_2019} for a sample of galaxies overlapping the Cepheid sample of \citet{riess_2016}. \citet{chen_2024} have analyzed the Pantheon+ \sne sample that forms the basis of the \citet{riess_2022} study, and illustrate that the published range of choices for values of M$_B$ result in variations in the mean \ho values ranging from 2\% to 7\% (i.e.,  up to 5 \hounits.) A critical issue remains the small sample size of nearby \sne available for absolute calibration in  the extragalactic distance scale \citep{riess_2022}. Furthermore, different groups employing varying \sn light-curve fitting techniques have yet to achieve convergence in their luminosity measurements \citep[e.g.,][and references therein]{vincenzi_2025}. Resolving the absolute magnitude calibration for \sne is essential for achieving high-accuracy \ho measurements.

\rm

\hst has now served for more than 30 years as the primary, space-based instrument for the discovery and measurement  of extragalactic Cepheid variables, and the local determination of \ho (e.g., the Hubble Key Project \citep{freedman_2001}, and \shoes \citep{riess_2022}). With its high resolution and sensitivity at optical wavelengths (where the amplitudes of Cepheids are large), \hst is an ideal instrument for {\it discovery} of extragalactic Cepheids. In contrast, \jwst is the best-equipped  telescope for improving the {\it accuracy} of measurements of \ho, due to  its greater sensitivity at long wavelengths (where Cepheid amplitudes are generally lower) and its even higher spatial resolution relative to \hst. The long-wavelength capability of \jwst also makes it an ideal facility for the study of other distance indicators; for example, the redder TRGB and JAGB (J-Region Asymptotic Giant Branch) populations. Additionally, in the near- and mid-infrared, interstellar extinction is significantly lower than in the optical  (e.g., A$_J$ and A$_{[4.4]}$ are smaller by factors of 4$\times$  and 20$\times$, respectively, relative to the visual extinction, A$_V$; and factors of 2$\times$  and 10$\times$ lower relative to the I-band extinction, A$_I$ \citep{cardelli_1989, indebetouw_2005}. Moreover, the science performance of \jwst has exceeded initial estimates of its sensitivity, stability, image quality, as well as spectral range \citep{rigby_2022}. \nircam (F115W) imaging from \jwst \citep{rieke_2023} has a sampling resolution four times higher than \hst $WFC3$ (F160W),  with a FWHM of 0.04~arcsec  on \jwst, versus 0.15~arcsec on \hst.  Of some potential concern, at red wavelengths, red giant and bright asymptotic giant branch stars can impact the photometry of the Cepheids (due to crowding and blending effects), exacerbating these effects in the red, compared to optical wavelengths. {\it Importantly, with four times better resolution than \hst at these near-infrared wavelengths, crowding effects are decreased by more than an order of magnitude in flux when using \jwst}. 

In this paper, we present our results from a new long-term Chicago-Carnegie Hubble Program (CCHP) using the James Webb Space Telescope (\jwst). The aim of the program is to reduce the systematic uncertainties in the local extragalactic distance scale, and ultimately to provide a  robust measurement of \ho. Our goals are: (1)  to use  three independent stellar distance indicators (Cepheids, the TRGB, and JAGB stars) to obtain three high-precision \jwst-only  distances to  each calibrating galaxy, thereby reducing the overall systematic distance uncertainties, (2) to make use of the high resolution of \jwst to understand and reduce the possible effects of crowding and blending of Cepheids  in previous \hst photometry, (3) to   improve the corrections for dust, and (4) to improve constraints on the effects of metallicity on the Cepheid Leavitt law. Preliminary results from this program  have been published in \citep{lee_2024a, owens_2025a,   hoyt_2025a, freedman_madore_2023a, freedman_madore_2023b}.  Details of the measurements   for the TRGB and JAGB results discussed in this paper
are presented in two companion papers: \citet{lee_2024c}, \citet{hoyt_2025b} [hereafter L25 and  H25, respectively]. A third paper on the Cepheid distance scale \citep{owens_2025b} is in preparation.
We note also that recent \jwst results have also been reported by \citet{yuan_2022,riess_2023, li_2024,riess_2024a,riess_2024b, li_2025}. \rm

\begin{figure*} 
\includegraphics[width=1.1\textwidth]{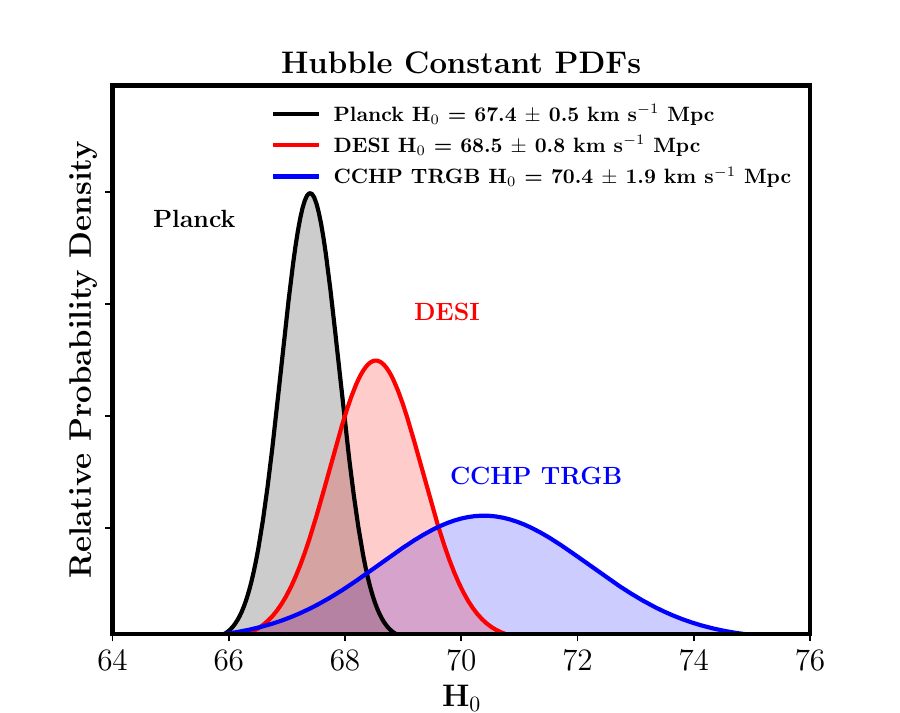}
 \caption{Relative probability densities for \ho values from Planck \citep{planck_2018}, DESI \citep{DESI_2024} and the \cchp TRGB measurement from this study.  Our current best estimate of \ho is  70.39 $\pm$ 1.22 (stat) $\pm$  1.33 (sys) $\pm$ 0.70 ($\sigma_{SN}$) \hounits. The results are all consistent, to within their uncertainties.
\label{fig:Ho_PDFs}}
\end{figure*}

The outline of this paper is as follows: In \S \ref{sec:overview} we provide an overview of the CCHP. We then describe the galaxy sample and analysis in \S \ref{sec:data}, our blinding procedure in \S \ref{sec:blinding}, and our calibration in \S \ref{sec:anchor}. In \S \ref{sec:distances}  we provide a description of the  two \rm methods used in this paper:  the TRGB and the JAGB. In \S \ref{sec:jwstdistances} we summarize our steps for the measurement of our new distances. In \S \ref{sec:comparison} we compare the distances obtained for the  two methods, followed by a comparison of previous TRGB measurements in \S \ref{sec:trgbprevcompare}.
A description of the calibration of \sne and determination of \ho is given in \S \ref{sec:h0}. 
In \S \ref{sec:sncalsys} we discuss the uncertainty that results from the small number of \sn calibrating galaxies. 
We discuss the current status of the Hubble tension in \S\ref{sec:tension}. Section \ref{sec:errors} presents a discussion of the overall uncertainties and an error budget. In \S \ref{sec:previous} we compare our results with previously published data. 
Section \ref{sec:future} provides a description of future prospects. Finally, in \S \ref{sec:conclusions} we present our summary and conclusions. We anticipate the main results of this study in Figure \ref{fig:Ho_PDFs}. 
\par\noindent

\medskip\medskip\medskip

\section{The Chicago-Carnegie Hubble Program: An Overview}
\label{sec:overview}

The CCHP was designed as a follow-up to the \hst Key Project, (originally named the Carnegie Hubble Program, CHP). Its primary goal was to decrease the systematic uncertainties in the measurement of \ho. Initially it began with a program designed to utilize the mid-infrared capability of the Spitzer Space Telescope, in anticipation of the parallax satellite \gaia. Our early infrared initiative provided 3.6 and 4.5 $\mu$m data for Cepheids in the Milky Way \citep{monson_2012}, the Large (LMC) and Small (SMC) Magellanic Clouds and the Local Group dwarf irregular galaxy, \ic \citep{scowcroft_2011, scowcroft_2013, scowcroft_2016a, scowcroft_2016b}. Based on a 3.6 $\mu$m distance to the LMC of 18.477 $\pm$ 0.033 mag, a recalibration of the \hst Key Project data \citep{freedman_2012} resulted in a value of \ho = 74.3 $\pm$ 2.1 (2.8\%) \hounits, reducing the formal uncertainty by a factor of three. 

In its second phase, the CCHP undertook a calibration of the Tip of the Red Giant Branch (TRGB) using the \hst {\it Advanced Camera for Surveys (ACS)} to derive distances to nearby galaxies in which Type Ia supernovae (\sne) had been discovered, and thus provided a calibration  that was independent from that based on Cepheids alone. TRGB distances were measured to 15 galaxies that were known hosts to 18 \sne. The halos of these galaxies were intentionally targeted to minimize the effects of dust, as well as to minimize contamination by younger and brighter disk asymptotic giant branch (AGB) stars. At the same time, crowding was also reduced by pre-selecting low surface brightness regions in the outer halo. The resulting distances were then used to tie into a  more distant sample of  \sne,  observed with high cadence and 9 filters (from the ultraviolet to the near infrared)  as part of the Carnegie Supernova Project (\csp) \cite{krisciunas_2017}. This effort  resulted in a value of \ho = 69.8 $\pm $ 0.6 (stat) $\pm$ 1.6 (sys) \hounits \citep{freedman_2019, freedman_2020, freedman_2021}. The results were little changed if instead the \sne catalog from the \shoes collaboration \cite{scolnic_2015} was adopted, giving \ho = 70.4 $\pm$ 1.4 \hounits \citep{freedman_2019}.

In the current and third phase of the CCHP, we  have undertaken a three-pronged approach to measuring \ho using Cycle 1 \jwst observations (JWST-GO-1995: P.I. W. L. Freedman). We observed 10 nearby galaxies that are hosts to 11 type \sne, and measured three independent distances  using Cepheids, the TRGB and \jagb stars.  In addition, we obtained observations of \ngc~4258, a galaxy  that is host to H$_2$O megamasers, and which provides a  geometric distance and our absolute calibration.   The program was designed to deal specifically with known systematic effects in the measurement of distances to nearby galaxies: (a) extinction and reddening by dust along the total line of sight to these objects, (b) metallicity effects, and (c) crowding/blending of stellar images. Simply getting more nearby galaxy distances (decreasing the statistical uncertainties) is insufficient to confirm or refute whether new physics beyond the standard cosmological model is required. At this time, systematic uncertainties are (and  have historically always been) the dominant component of the error budget. 

\section{The Galaxy Sample, Data Acquisition and Analysis}
\label{sec:data}

\begin{figure*} 
 \centering
\includegraphics[width=15.0cm]{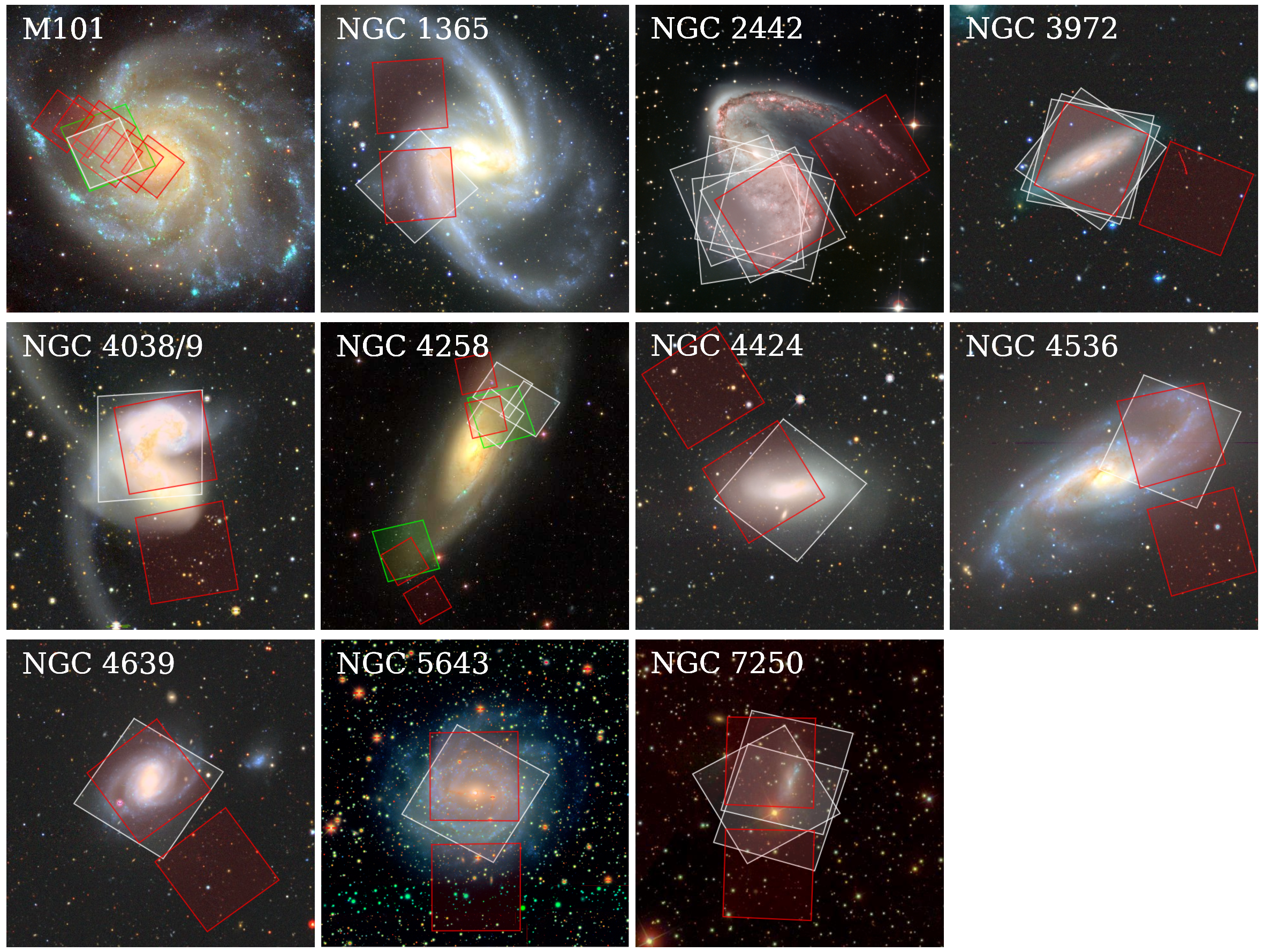}
 \caption{Images of the 11 galaxies observed as part of this program. North is up and east is to the left. The images have been obtained from the following public sources: SDSS: M101 and NGC 4258 ; DECaLS: NGC 1365, 3972, 4038, 4424, 4536, 4639, 7250 ;ESO: NGC 2442 ; and NOIRLab: NGC 5643. 
The red and white squares denote the footprint of JWST NIRCam and HST WFC3, respectively. The small green squares in M101 and NGC 4258 are ACS WFC.
}
\label{fig:galaxyimages}
\end{figure*}

The galaxies chosen for this program have (1) well-observed \sne with well-defined light curves and peak magnitudes, (2)  previously-discovered Cepheid variables \citep{freedman_2001,riess_2022}, and (3) are located at distances $\lesssim$ 23 Mpc, suitable for measuring accurate Cepheid, TRGB and JAGB (carbon-star) distances. All three of these methods are of high precision and can be independently used to calibrate \sne. Images of our 11 target galaxies are shown in Figure \ref{fig:galaxyimages}.

Observations were obtained using the \jwst Near-Infrared Camera (\nircam; \cite{rieke_2023}), with the  $F115W$ filter (comparable to ground-based $J$ band) and the F356W filter at 3.6$\mu$m. The observations were carried out  over a 13-month period of time from  November 2022 to January 2024. The  first observations were obtained for the galaxies \ngc~7250 and \ngc~4536  (\cite{lee_2024a, freedman_madore_2023a, freedman_madore_2023b, hoyt_2025a, owens_2025a}),  making use of the F444W filter at 4.4~$\mu$m, to both provide a metallicity test for Cepheids, and a long color baseline for the discovery and color discrimination of the JAGB stars. We then switched to F356W for the rest of the target sample, owing to its higher sensitivity and better spatial sampling. However, the F444W filter contains a CO bandhead that is sensitive to metallicity \citep{scowcroft_2016b}, and it is being used to carry out a test for metallicity effects in the two nearest galaxies, M101 and \ngc 4258 (at distances $\lesssim$7.5 Mpc) where the highest resolution can be achieved.

\nircam covers a total  area of 9.7 arcmin$^2$ field with a 44~arcsec gap between two 2.2 × 2.2 arcmin regions. Our target fields were chosen to optimize the inclusion of (1) the largest possible number of known Cepheids in the inner disk, (2)  portions of the extended outer disk to detect carbon stars, and (3) with a rotation angle optimized for the detection of halo red giants \citep{hoyt_2025a}.

A detailed description of our data reduction procedures can be found in \citet{jang_2025}, hereafter J25. Here we provide a brief overview.  The images\footnote{We obtained the stage 2 \texttt{\_cal} images with jwst\_1149.pmap from the MAST archive.} were processed primarily using the \nircam module in the software package DOLPHOT, updated for \jwst data analysis \citep{weisz_2023, weisz_2024}. We found a significant improvement in the point-spread modeling for this updated version of DOLPHOT\footnote{DOLPHOT 2.0 NIRCAM module sources (updated 4 Feb 2024)}, resulting in considerably better source identification and image subtraction using a so-called `warmstart' mode. In addition, we are carrying out a parallel and independent analysis of the frames using DAOPHOT \citep{stetson_1987} in order to provide a quantitative constraint on photometric errors that might arise due to differences in point-spread-function fitting and sky-subtraction approaches in crowded fields.  
In the case of one of our outer fields in the anchor galaxy, \ngc 4258, upon which our current calibration is based, we find excellent agreement between the DOLPHOT and DAOPHOT reductions at a level of 0.002 mag ($<$0.1\%) with a total $rms$ of 0.02 mag down to 3 mag below the TRGB \citep{hoyt_2025b}. The comparisons for our other galaxies are ongoing. We also tested the automatic aperture corrections (Apcor = 1) provided by DOLPHOT by comparing the overlapping regions of dithered images, as well as by carrying out comparisons with manual/visual determinations of the aperture corrections. These tests confirmed that the  automatic aperture corrections were good to a level of $<$0.02 mag. Finally, {\it  all of our initial data analysis was carried out using a blinding procedure}, as described in \S \ref{sec:blinding} below.

\begin{figure*} 
 \centering
\includegraphics[width=16.0cm]{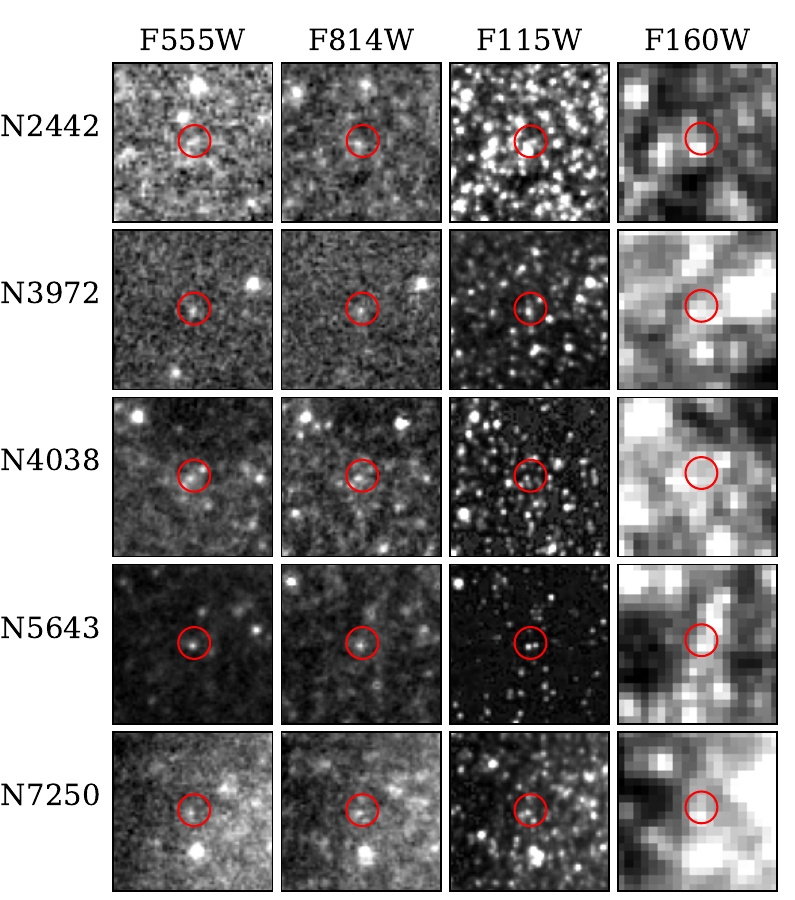}
 \caption{Examples of Cepheids in \ngc 2442, \ngc 3972, \ngc 4038, \ngc 5643 and \ngc 7250 shown in \hst filters $F555W, ~F814W$ and $F160W$, as well as the \jwst filter $F115W$. These cutouts are made from drizzled images, with pixel scale equal to 0.035” in $F555W, F814W$, and $F115W$, and equal to 0.10” in $F160W$. Thus, for consistency in comparison, these pixel scales are comparable to the average pixel size of the detectors used with each filter. In each cutout image, the maximum pixel value is set to the brightness at the center of the Cepheid point spread function. With this scaling, any white pixel on the cutout images is equal to or greater in brightness than the target Cepheid itself. From this figure, the higher resolution of the near-infrared \jwst ($F115W$) images can be seen relative to those of \hst ($F160W$). In addition, it can be seen that the optical HST images ($F555W$ and $F814W$) are of higher resolution than the $F160W$ images owing both to the better resolution and less contamination from red giants and asymptotic giant branch stars. 
}
\label{fig:Cephcutouts}
\end{figure*}

We note that for our ongoing Cepheid analysis, we are making use of archival F350LP, F555W and F814W \hst data in combination with the F115W \jwst near-infrared data. The three wavelengths (F555W, F814W and F115W) jointly allow a more precise measurement of the extinction.  However, the optical data (F555W and F814W) are more heavily affected by crowding (see Figure \ref{fig:Cephcutouts}).
In contrast, for the TRGB and JAGB analyses presented in this paper, we used the \nircam F115W and the F356W and/or F444W data, and did not use any optical \hst data. The source coordinates were obtained from the F115W images, and  used in `warmstart' mode for the longer wavelength data (for more details see H25, L25 and J25). 

\begin{figure*} 
 \centering
\includegraphics[width=10.0cm]{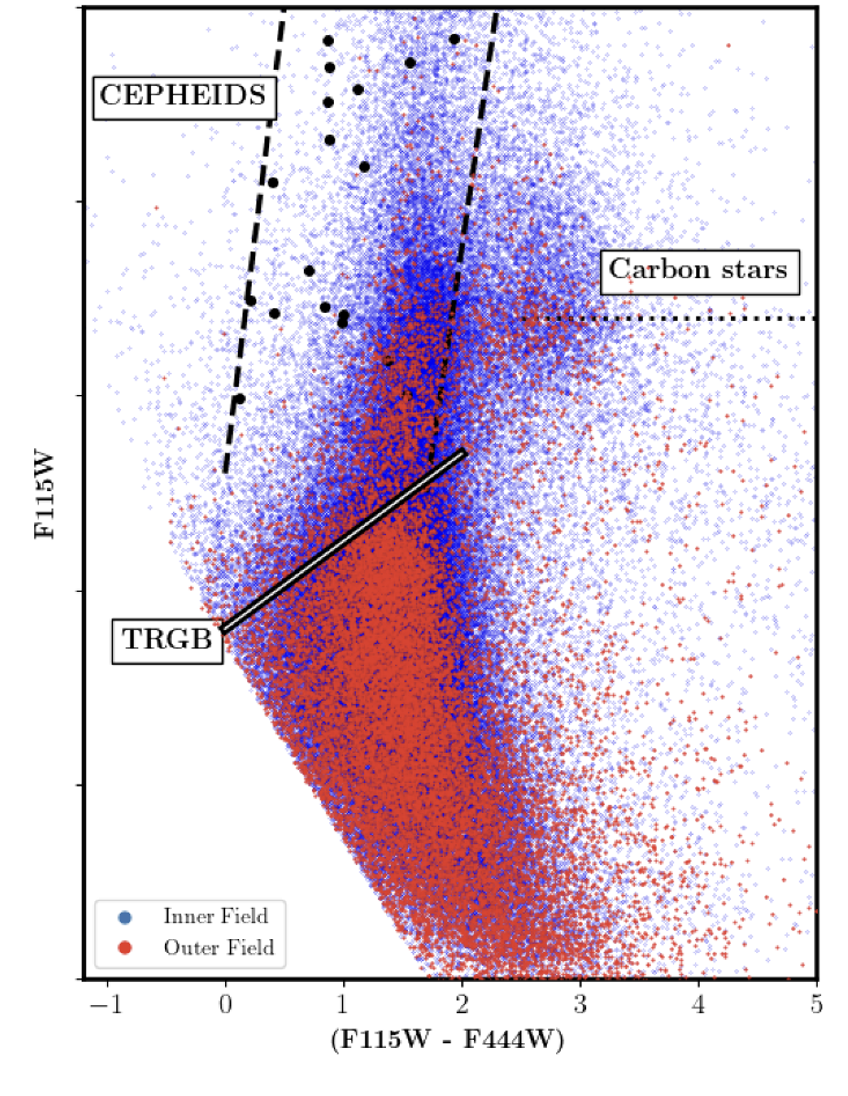}
 \caption{F115W versus (F115W-F444W) color-magnitude diagram for the galaxy \ngc 7250. The positions of all three distance indicators used in our program are identified in the plot. Schematically, the dashed slanted black lines indicate the approximate blue and red edges of the Cepheid instability strip; the white sloped line indicates the position of the TRGB; and the dotted line indicates the peak of the carbon/JAGB stellar luminosity distribution.  Red points are stars in the outer field of the galaxy; blue points are stars in the inner field of the galaxy.}
\label{fig:CMD}
\end{figure*}

We show in Figure \ref{fig:CMD} a color-magnitude (F115W versus F115W-F444W) diagram with the locations of all three of the stellar distance indicators indicated. The Cepheids are the brightest objects, followed by the redder JAGB stars and then the TRGB stars. The power of this program can be seen at a glance: the same detector, pixel scale, point spread functions, photometric reduction packages, and calibration are being applied to all three methods, simultaneously. A deliberate placing and orientation of the two \nircam detectors has allowed the TRGB to be measured in the halo, the Cepheids in the inner disk, and the JAGB stars in the outer disk.

\begin{deluxetable*}{lcccccc}
\tablecaption{CCHP Galaxy Sample}\label{tab:jwstgalaxies}
\tablehead{
\colhead{Galaxy} & 
\colhead{SN Name}& 
\colhead{Morphological Type\tablenotemark{a}} & 
\colhead{$<$O/H]$>$ \tablenotemark{b}}& 
}
\startdata
M101  & SN 2011fe &   SAB(rs)cd & ~0.10 dex\\
NGC 1365  &   SN 2012fr & SB(s)b & -0.14 dex\\
NGC 2442  &  SN 2015F & SAB(s)bc pec & ~0.00 dex\\
NGC 3972 &  SN 2011by  & SA(s)bc & ~0.03 dex\\
NGC 4038 &  SN 2007sr & SB(s)m pec & ~0.03 dex \\
NGC 4424  & SN 2012cg & SB(s)a: & ~0.06 dex\\
NGC 4536 &  SN 1981B &  SAB(rs)bc &-0.15 dex \\
NGC 4639 &  SN 1990N & SAB(rs)bc & -0.01 dex \\
NGC 5643 &   SN 2013aa, SN 2017cbv 	& SAB(rs)c & ~0.13 dex\\
NGC 7250&  SN 2013dy & Sdm:  & -0.28 dex\\
\hline
\enddata
\tablerefs{(a) NED (NASA Extragalactic Database) }
\tablerefs{(b) from \cite{riess_2022}, Table 3; Solar Value of 12+ log[O/H] = 8.69 dex, based on \cite{asplund_2009}.  }
\end{deluxetable*}

\section{Our Blinding Procedure}
\label{sec:blinding}

Our initial goal was to undertake all of the data analysis without knowledge of the true zero-point calibration, starting from the receipt of the raw data frames, through to the final analysis and determination of \ho for the three independent methods. In practice, the blinding experiment was (completely) successful only for the JAGB analysis, as described below. 

To set up the blinding procedure, the frames were processed using DOLPHOT, as described above. Random numbers were then generated and added separately to each of the photometry catalogs based on the raw data frames. The same random offset was added to all passbands in each galaxy, preserving the colors. The blinding was such that during the entire year and a half of the photometric analysis, no one in the group had any knowledge of what the true distances or the value of \ho might be. 

One week before unblinding, one team member was given access to the unblinded photometry. This step was taken to test for any potential catastrophic errors somewhere in the blinded analysis. None were found. As part of this effort, an independent analysis, using independent software (fitting of the PL relations, measurement of the TRGB and JAGB luminosity functions) was initiated. 

For the TRGB, spatial cuts of the data, selection of TRGB stars, slopes of the TRGB, rectification, Sobel edge detection and measurement of the TRGB were made with the random offsets applied to the individual data frames. Finally, TRGB  distances, with arbitrary zero-points, were measured with respect to \ngc 4258 (the latter also having a random offset). Similarly, in the case of the JAGB stars, all analysis through spatial cuts (and decisions about whether there was convergence in the spatial cuts),  measurement of the luminosity functions, and measurement of distances were carried out with blinded, arbitrary zero points.

In the case of the Cepheid distance scale, subsequent tests after unblinding revealed an issue in the data pipeline and in the photometric reductions of the \hst data. A re-analysis of these data is on-going. 

Finally, the entire group met in person on March 13, 2024, and we unblinded the photometry with everyone present, with five of the six co-authors seeing the unblinded results for the first time. We applied the calibration based on \ngc 4258 (which until that time had also had an arbitrary zero point) and compared the distances that had been obtained independently using the three different methods (as described in \S \ref{sec:comparison} below).  At our unblinding meeting, we applied the new local calibrating galaxy distances to obtain distances to the Carnegie Supernova Program and Pantheon+ \sn data (\S \ref{sec:hubdiag}), and independently determined preliminary values of \ho based on the two \sn samples.  At the time of unblinding the agreement among the three distance indicators was extremely good.

No further updates were made to the JAGB analysis following unblinding. We can consider the JAGB distances obtained in a completely blinded fashion, and this rung of the blinding experiment a success from start to finish. In the case of the TRGB, almost all aspects of the analysis remained unchanged after unblinding, except for the sample-wide error estimation and the spatial selection for four of the thirteen fields, all of which had the four largest uncertainties in the blinded analysis. Closer inspection of the color-magnitude diagrams in all of these cases revealed an underpopulated tip that was filled out with the post-unblinding larger spatial selection.  (For further details see H25.) The sense of the changes was to decrease the distances in these cases, with a subsequent increase to the value of \ho. These differences were within one-sigma of the blinded results.

\section{NGC 4258 and Photometric Zero-point Calibration}
\label{sec:anchor}

Our zero-point calibration for the \jwst data is based on the geometric distance to the galaxy, \ngc 4258. For ground and \hst distance scale measurements, the LMC  and the Milky Way also serve as calibrators. Unfortunately, the Milky Way Cepheids, TRGB and JAGB stars are too bright to be observed with \jwst.  \ngc 4258 is a nearby,  highly-inclined  spiral galaxy
located at a distance of 7.6 Mpc. A geometric distance to the galaxy  has been measured via its H$_2$O masers, which are orbiting within an  accretion disk inclined at $\sim$ 72$^\circ$,  surrounding a supermassive black hole \citep[see][]{humphreys_2013, reid_2019} and megamaser. The most recent geometric distance to \ngc 4258 is 29.397 $\pm$ 0.024 (stat) $\pm$ 0.022 [sys] mag \citep{reid_2019}, a 1.5\% measurement.

There are several advantages to a calibration  based on \ngc 4258. Importantly, it can be applied consistently across all three of the methods (Cepheids, TRGB and JAGB stars).  Unlike the case of the nearby Milky Way Cepheids and those in the LMC, crowding and blending effects for \ngc 4258 are comparable to those of the more distant galaxies, and the range of magnitudes  for the  Cepheids, TRGB and JAGB stars in both the calibrator and target galaxies allows them to be observed with the same telescope and instrument. Furthermore, by adopting the distance to \ngc 4258, it requires simply \jwst flight magnitudes alone, without recourse to ground-based tie-ins.  Finally, the metallicity of \ngc 4258 is comparable to the mean metallicity of the Cepheid fields used in our target galaxies. 

Inspecting Table \ref{tab:jwstgalaxies}, our \jwst galaxy sample has a range in average  metallicity going from -0.28 $<$ [O/H] $<$ 0.13 dex, with a mean of -0.023 dex. This value is comparable to that of the R22 sample of 37 galaxies (their Table 3), which has a mean metallicity of -0.037 dex. \ngc 4258 has a metallicity of [O/H] = -0.10 dex. The difference in average metallicity for our sample with respect to  \ngc 4258 is -0.077 dex. For a value of $\gamma$ = -0.2 mag/dex (as adopted by R22), this translates to a Cepheid metallicity correction in the mean of less than  1\% (i.e., -0.7\%) to \ho.

An external check of  the \ngc4258 calibration can be obtained by comparing previous results for the case where the determination of \ho was based on several anchors (\ngc~4258, Milky Way, LMC, and SMC) rather than \ngc 4258 alone. In the case of the TRGB,  applying distances for these four anchors resulted in (internally) good agreement:  \citet{freedman_2021} found values of 69.7 $\pm$ 1.0 (stat) $\pm$ 2.0 [sys] for \ngc~4258, 69.3 $\pm$0.8 (stat) $\pm$ 3.5 [sys] for the Milky Way,  69.9 $\pm$ 0.5 (stat) $\pm$ 1.6 [sys] for the LMC and  69.5 $\pm$ 1.0 (stat) $\pm$ 1.7 [sys] for the SMC calibrations, respectively. In Appendix \ref{App:AppendixA}, we undertake a preliminary investigation of the calibration of the JAGB method based on the Milky Way, LMC \& SMC, and find very good agreement in the zero-point calibration with that for \ngc 4258.
Similarly, good internal agreement for the Cepheid calibration was also found by R22 with  values of 72.51 $\pm$ 1.54 for \ngc 4258, 73.02 $\pm$ 1.19 for the Milky Way, and 73.59 $\pm$1.36 for the LMC calibrations, respectively. Thus,  having \ngc~4258 as the sole calibrator cannot explain the reason for the  difference in \ho between 69 and 73: the internal agreement amongst the anchors is excellent for  all of the methods. 

\medskip\medskip\medskip\medskip\medskip\medskip\medskip\medskip

\section{Measurement of Distances}
\label{sec:distances}

In this section we give a brief overview of the two methods used in this paper, and discuss their advantages and disadvantages. There is no single perfect method. The weaknesses of each method serve to underscore the importance of having independent methods for constraining overall systematic effects in the local distance scale.

\subsection{TRGB}
\label{sec:trgb}

The use of red giant branch stars as distance indicators goes back over a century to Harlow Shapley, who measured the brightest stars in globular clusters, as one of his techniques for determining the size of the Milky Way galaxy \citep{shapley_1918}. Beginning in the 1980s, the availability of CCD detectors, with red sensitivity and high quantum efficiency, resulted in a resurgence of interest in these stars, this time in their application to the extragalactic distance scale \citep{mould_kristian_1986, da_costa_armandroff_1990, lee_1993}. 

The TRGB now provides one of the most precise and accurate means of measuring distances to nearby galaxies, comparable to the Cepheid Leavitt law. In practice, the observed color-magnitude diagrams of old red giant branch stars display a distinct edge/discontinuity in the red giant branch (RGB) luminosity function (LF). This astrophysical discontinuity corresponds to the core helium-flash luminosity at the end phase of RGB evolution for low-mass stars. Measurement of  the TRGB edge has repeatedly been shown to be an excellent standard candle in the $I$ band \citep{lee_1993, rizzi_2007, salaris_2002, madore_2009, freedman_2019, jang_2021}, and it is a standardizable candle in the near infrared (\cite{dalcanton_2012, wu_2014, madore_2018, durbin_2020, hoyt_2025a}).  Details of the method can be found in a number of reviews \cite[e.g.,][]{madore_freedman_1999,freedman_madore_2010,beaton_2019b,madore_freedman_2023}.

A strength of the TRGB method is that the underlying theory, for why it is an excellent (empirical) standard candle, is well understood \citep{salaris_1997, Salaris_2005, bildsten_2012,kippenhahn_2013, serenelli_2017, saltas_tognelli_2022}. Using a grid of $\sim$3 million TRGB models covering a wide range of metallicities and initial masses, \citet{saltas_tognelli_2022} recently examined the effects of chemical composition, opacity, nuclear reactions, and convection; and compared these results to other published models. Based on a Monte Carlo analysis, they found a maximum theoretical uncertainty of only $\sim$1.6\%, which is primarily due to radiative opacity systematics. The TRGB is currently the only method for distance determination with a well-understood theoretical basis. 

After leaving the main sequence, low-mass stars with masses $M\lesssim 2 M_{\odot}$ develop a degenerate helium core. They ascend the red giant branch, with their luminosity being powered by a hydrogen-burning shell surrounding the core. The freshly formed helium adds to the core mass  until it reaches a  value of about 0.5 M$_{\odot}$, {\it independent of the initial mass of the star}. When the temperature reaches about 10$^8$K in the isothermal core, it enables the triple-alpha process (helium burning) to commence. Because the degenerate core cannot expand, a thermonuclear runaway ensues (the core helium flash),  injecting  energy  that eventually lifts the electron degeneracy. The star then rapidly evolves away from the red giant branch and descends in luminosity onto the  horizontal branch or the red clump, where it undergoes stable core helium burning at a lower-luminosity.

The advantages of the TRGB method are: (1) The discontinuity in the observed TRGB luminosity function is empirically simple to identify and measure. (2) The physical mechanism for the TRGB (the core helium flash) is well understood. (3) If applied in the outer regions of galaxies, where the surface brightness of the galaxy is low, then the overlapping (i.e., crowding or blending effects) of the stellar point spread functions is minimal. (4) In the outskirts of galaxies, the effects of dust are small \citep[e.g.,][]{menard_2010}. 
(5) The metallicity of a star on the red giant branch relates directly to its observed color  \citep[e.g.,][]{da_costa_armandroff_1990, carretta_bragaglia_1998}, and can be easily calibrated and corrected for, if required. (6) For a measured TRGB slope in a given passband, the slope of the giant branch luminosity in a different passband is uniquely defined mathematically   \citep{madore_freedman_2020}.

The disadvantages of the TRGB method are: (1) Care must be taken to ensure that the regions in the target and calibrating galaxies are comparable in terms of surface brightness and line-of-sight column density of dust. (2) Target field placement is critical. Too close in and crowding/blending/reddening become issues. Too far out and the number density of stars is too low for a precise measurement of the tip. (3) Occasionally adjacent peaks of comparable strength can be found in poorly sampled or heavily contaminated luminosity functions, complicating (or making impossible) an accurate measurement of the true tip. Cases (2) and (3) can be remedied with the acquisition of augmented sample sizes.

\medskip\medskip\medskip\medskip\medskip

\subsection{JAGB Stars}
\label{sec:jagb}

Although relatively new in the context of the  extragalactic distance scale, J-region AGB (JAGB) stars were, in fact, first identified in the LMC nearly a quarter of a century ago, as a distinct class of objects \citep{nikolaev_weinberg_2000, weinberg_nikolaev_2001},  when they were used to determine the back-to-front geometry of the LMC. The JAGB method is now one of the most promising methods for measuring the distances to nearby galaxies. (1) These thermally-pulsating AGB carbon stars have a nearly constant luminosity in the near-infrared J band (at 1.2 $\mu$m), and (2) they have a low intrinsic dispersion  of only $\pm$0.2 mag (\cite{nikolaev_weinberg_2000, madore_freedman_2020}). Moreover, (3) JAGB stars can be easily and unambiguously identified on the basis of their near-infrared colors alone (without the need for spectroscopy or narrow-band photometry), thereby being readily distinguished from both the (bluer) O-rich AGB stars, as well as (even redder) ``extreme'' carbon stars. 

Extending the JAGB method, \cite{freedman_madore_2020} used these stars to measure the distances to additional nearby galaxies within and beyond the Local Group, out to 27 Mpc. In a sample of 14 galaxies, they found excellent agreement of the JAGB distances with published TRGB distances to those same galaxies, where the combined scatter (including potential effects of metallicity and star formation history) amounted to only $\pm$4\%.  Simultaneously, \cite{ripoche_2020} investigated the Magellanic Clouds and the Milky Way. A number of additional extensive tests of this method have recently been carried out by Lee and collaborators \cite{lee_2021, lee_2022, lee_2023a, lee_2024a, lee_2024b} as well as \cite{parada_2021}, \cite{zgirski_2021}, and \cite{magnus_2024} in several nearby galaxies, confirming the excellent agreement with distances measured with the TRGB and Cepheid distance scales, and again indicating that any potentially confounding effects of metallicity variations and star formation history must be contained within that small scatter, and therefore must be relatively small. 

Two astrophysical  effects are known to account for the small spread in the  luminosity of carbon stars. Carbon stars are formed during the thermally pulsing  phase for AGB stars when carbon is convectively dredged up to the surface of the star (\cite{iben_renzini_1983, herwig_2013, habing_olofsson_2004, marigo_2017,  salaris_2014, pastorelli_2020}).  (a) The upper limit to the luminosity results from the fact that younger, more massive (hotter) AGB stars burn their carbon at the bottom of the convective envelope before it can reach the surface of the star \cite{boothroyd_1993}. (b) The lower limit to the luminosity of carbon stars results from the fact that for the oldest, less massive AGB stars, there is no third (deeper) dredge-up phase. Thus, carbon stars emerge only in a well-defined intermediate mass range: where thermal pulses are effective dredging up carbon to the surface, but only when the carbon survives being burned at the lower levels before being mixed into the outer envelope.

The advantages of the JAGB method are many. (1) JAGB stars are the dominant population of the reddest stars found in galaxies. They are easily identified on the basis of their near-infrared colors. (2) The color-selected $J$-band luminosity function is centrally peaked with a low dispersion. (3) JAGB stars are, in the mean, about one magnitude brighter than the brightest TRGB stars. (4) JAGB stars are found in all galaxies that have an intermediate-age population, and therefore the method can be applied to a wide range of galaxy types. (5) In the infrared, the total  reddening along the line of sight (including Milky Way foreground extinction, host galaxy extinction, and any dust reddening generated by the carbon stars themselves) drops by about a factor of four in going from the optical ($V$) to the near infrared ($J$). (6) Another major advantage is that JAGB stars can be found in the outer disks of galaxies where crowding and reddening effects are significantly reduced compared to the inner disks where Cepheids reside. And, (7) unlike the use of period-luminosity relations for variable stars (e.g., Cepheids or Miras), which need to be monitored over multiple cycles, JAGB  stars require only a single-epoch observation. 

The disadvantages of the JAGB method are (1) Although minimized in the $J$ band, reddening variations within the host galaxy remain a source of uncertainty for the JAGB population. (2) Care must be taken to compare regions of comparable dust and metallicity levels in the calibrating and target galaxies. And (3) the comparative novelty of this method requires further study to quantify potential effects that might result, for example, from differences in star formation histories, or the existence of different population sub-types.

\medskip\medskip\medskip\medskip\medskip\medskip\medskip\medskip

\section{New Distances from JWST Observations}
\label{sec:jwstdistances}

As noted in \S\ref{sec:data}, our observing program was configured to optimize simultaneous observations for all three of our methods\footnote{The {\it JWST} data for our program are available from the Barbara A. Mikulski Archive for Space Telescopes (MAST) Portal, Proposal ID 1995.}. This goal led us to choose  $F115W$ as our primary filter. First, JAGB stars have nearly constant luminosities in the $F115W$ ($J$) band. Second, $F115W$ observations provide an additional waveband beyond the \hst $F555W, F814W$ and $F160W$ filters, to improve the extinction correction for Cepheids. Third, extinction effects are small at 1.15 $\mu$m, which is an advantage for the TRGB method relative to the $I$ band (where extinction effects are about a factor of two larger).   We note that the disadvantage of the $J$-band for the TRGB method is that there is a steeper color dependence than for the $I$ band where the TRGB is relatively flat \citep[e.g.,][]{lee_1993}. Nonetheless, the TRGB is still standardizable in the near-infrared \citep[e.g.,][]{dalcanton_2012, madore_2018, hoyt_2018, durbin_2020,  newman_2024b, hoyt_2025a}. Solving for a slope makes the method more analogous to that of measuring a period-luminosity relation for Cepheids.

As also noted in \S\ref{sec:data}, our original choice for our second filter was motivated to allow us to provide constraints on the metallicity sensitivity of the Leavitt law. At 4.5 $\mu$m there is a  metallicity dependence of the Leavitt law as a result of a CO-band head \citep{scowcroft_2011}. The $F444W$ filter would also provide a broad color baseline for the JAGB method.  However, the relatively inferior resolution of the $F444W$ imaging data prompted us to switch to the $F356W$ filter, which provides both higher resolution and higher sensitivity. 

\subsection{$JWST$ TRGB Measurements}
\label{sec:jwsttrgb}

An important aspect of accurately measuring  TRGB distances is in selecting for an old population. Doing so requires an optimal balance between avoiding regions of the disk where younger populations reside, and not being so far out in the halo so that the number density of stars is too small  (for a fixed-sized detector footprint on the sky).  As described in H25, we used the unresolved blue light to trace regions dominated by young disk populations, masking those regions bright in blue flux and with noisy, unclear TRGBs.

\begin{figure*} 
 \centering
\includegraphics[width=15.0cm]
{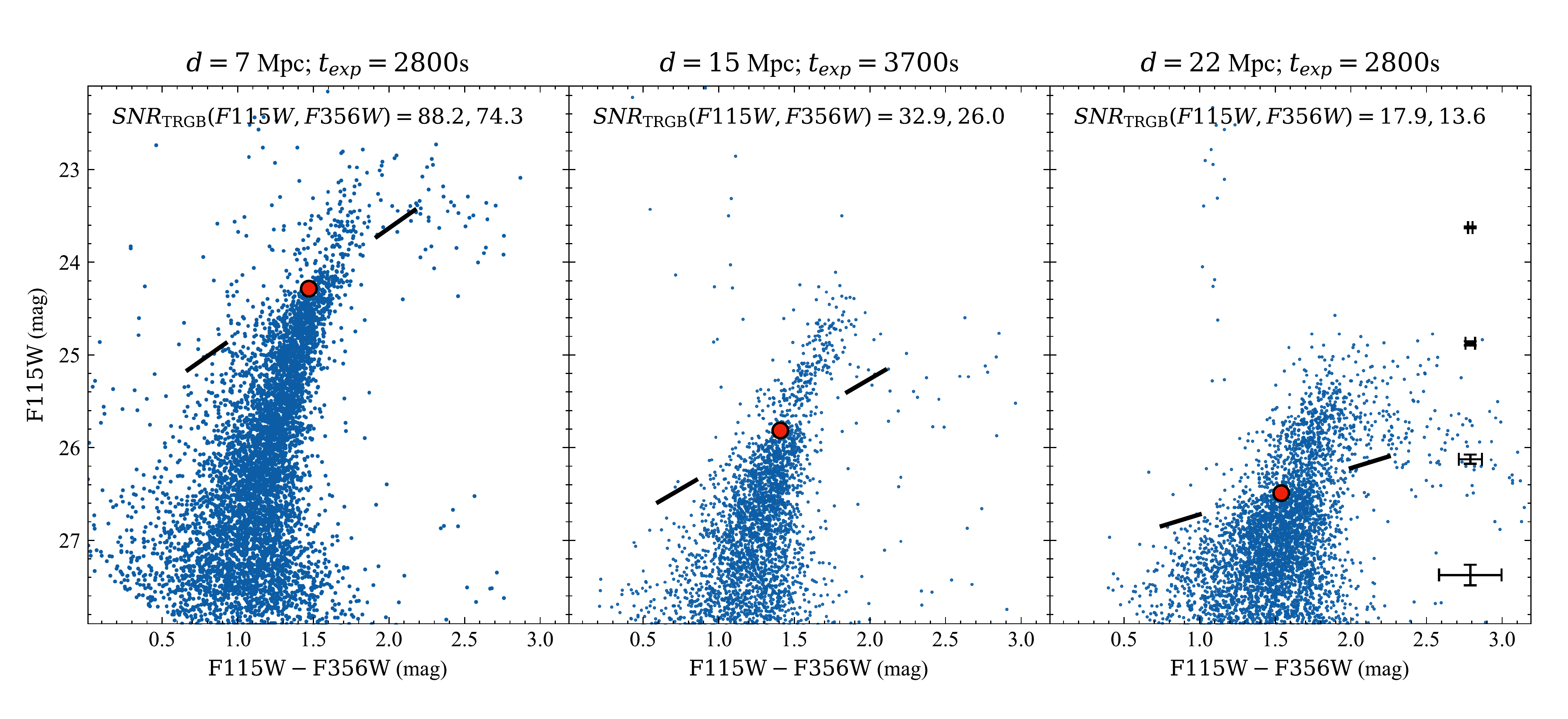}
 \caption{$F115W$ versus ($F115W$ - $F356W$) color-magnitude diagrams for \ngc 4258, \ngc 4424 and \ngc 4039. The distances, exposure times and signal-to-noise ratios for stars at the TRGB are given at the top of the respective plots. The red dot in each plot indicates the mean TRGB magnitude and color. A selection of mean error bars, plotted vertically as a function of the F115W magnitude, are shown at the right of the last plot. The galaxies shown span a range of distances from 7 Mpc to 22 Mpc, with a range of 88 to 18 in the signal-to-noise ratio at the TRGB.  }
\label{fig:jwsttrgb}
\end{figure*}

In Figure \ref{fig:jwsttrgb}, we show examples of $F115W$ versus ($F115W$ - $F356W$) color-magnitude diagrams for three of our program galaxies, \ngc 4258, \ngc 4424 and \ngc 4039 from H25. The position of the TRGB is shown, and is (both qualitatively and quantitatively) easily identified. As can be seen, the TRGB feature in the infrared exhibits the expected  upward trend with increasing photometric color. This dependence is due to a combination of metallicity and, to a much lesser extent, age effects \citep[e.g.,][]{mcquinn_2019}. 
To measure the TRGB in each \sn host, the gradient in the magnitude direction was computed from a smoothed Hess diagram of the selected RGB stars, and the centroid of the dominant contour defined the TRGB in each field. The different fields show some variation in slope, as seen here, but overall there is a consistent single, universal slope plus dispersion, all of which is propagated into the TRGB distance error budget \citep[see][for further details]{hoyt_2025b}.

\subsection{$HST$ TRGB Measurements}
\label{sec:newTRGB}

 Archival \jwst data from the \shoes Cycle 1 program (GO-1685 , PI: A. Riess) offer an important constraint on our TRGB calibration, given that they were taken with different filters that are complementary to our own observations. The galaxies observed include \ngc 1448, \ngc 1559, \ngc 4258, \ngc 5584 and \ngc 5643. The data have been reduced in a similar fashion to those described in our previous $F814W$ optical observations \citep[e.g.,][]{freedman_2019, hoyt_2019, jang_2021}. Two members of our CCHP group reduced the data independently in blind fashion, finding agreement in the average at the 0.015 mag (0.7\%) level.
 
\begin{figure}
\centering
\includegraphics[width=15.0cm, angle=-0]
{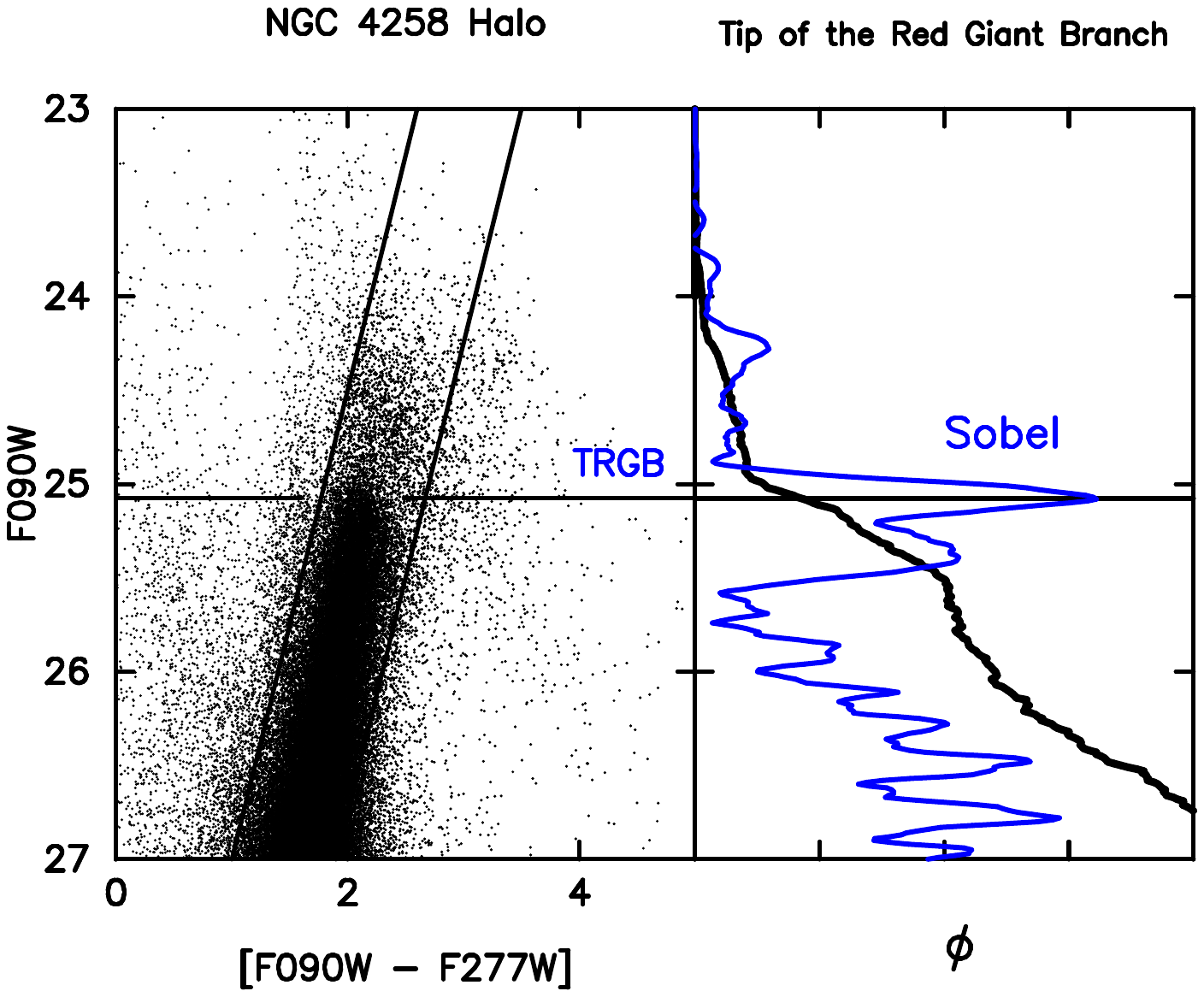}
\caption{Left panel: Archival \jwst \nircam Color-Magnitude diagram (CMD) of the halo of \ngc 4258. Left panel shows the redward-slanting red giant branch terminating at a peak brightness of $F090W$ = 25.08~mag. The TRGB is marked by the horizontal black lines corresponding to the maximum response of the Sobel edge-detector shown in blue in the right panel. The smoothed black line in the right panel is the marginalized $I$-band luminosity function of the RGB stars found to fall between the two thin upward-slanting black lines in the CMD to the left.
\label{fig:newTRGB4258}}
\end{figure}

 The data for \ngc 4258 were used to measure the $F090W$ zero point for the TRGB. In the left panel of Figure \ref{fig:newTRGB4258}, we show an $F090W$ versus ($F090W-F277W$) color-magnitude diagram for a halo region of  \ngc 4258. The position of the TRGB is shown by the solid black horizontal line at F090W = 25.08~mag. The right panel shows the marginalized and smoothed luminosity function in black, and the Sobel edge-detection filter output is shown in blue. Detection of the tip in the left panel, and its quantitative measurement in the right panel, is unambiguous. The luminosity function was smoothed by 0.1 mag before its digital first derivative was output by the Sobel filter. 

The \jwst $I$-band ($F090W$) data were derived from the sum of four 258-second \nircam exposures; the four $F277W$ equivalent exposures were taken in parallel, directed to the beam-split long-wavelength channel of \nircam. The Sobel response function has a measured $\pm$ 2-$\sigma$ width of $\pm$0.20~mag. Within the 4$\sigma$ full width of 0.4~mag there are 624 red giant branch stars contributing to the measurement.  We therefore  adopt a value of 0.4/$\sqrt(624) =\pm$0.02 mag for the statistical uncertainty. 

We adopt a  foreground $I$-band galactic line-of-sight extinction of $A_{F090W} = $ 0.02~mag for the halo of \ngc 4258, as derived by \citet{anand_2024a}. This gives a reddening-corrected $F090W$ magnitude of 25.06~mag for the TRGB in \ngc 4258. Adopting a geometric distance modulus of 29.397 $\pm$ 0.032 mag \citep{reid_2019} yields our zero-point calibration of M$_{F090W}$ =  -4.336 $\pm$ 0.02 (stat) $\pm$0.032 [sys]~mag. Note this value is on the Vega-Sirius system, the default magnitude system employed by the \jwst calibration program \citep{rieke_2023, gordon_2022}.
The measured distance moduli are  $\mu_{N1448} $ = 31.321 $\pm$ 0.049, $\mu_{N1559} $ = 31.491 $\pm$ 0.051, $\mu_{N5584} $ = 31.851 $\pm$ 0.053, and $\mu_{N5643} $ = 30.599 $\pm$ 0.057.
R24b measured TRGB distances of $\mu_{N1448} $ = 31.38 $\pm$ 0.07, $\mu_{N1559} $ = 31.50 $\pm$ 0.05, $\mu_{N5584} $ = 31.81 $\pm$ 0.09, and $\mu_{N5643} $ = 30.56 $\pm$ 0.06. These distance moduli agree very well, with a weighted mean difference amounting to only  -0.003 $\pm$ 0.038 mag (\cchp-\shoes).  Two additional galaxies from this program are not yet publicly available, \ngc 2525 and \ngc 3447.


\medskip\medskip\medskip\medskip\medskip\medskip\medskip\medskip\medskip

\subsubsection{Crosschecks of the TRGB F090W Zero Point}

Our zero-point value of M$_{F090W}$ =  -4.336 $\pm$ 0.02 (stat) $\pm$0.032 [sys]~mag compares favorably with the zero point of M$_{F090W}$ =  -4.32 $\pm$ 0.025 (stat)~mag published by \citet{newman_2024b}. They  averaged over six nearby galaxies observed with \jwst, using the same filter combinations as discussed here, but zeroed to previously published TRGB distances using an $I$-band $F814W$ calibration of M$_I$ =  -4.05~mag \citep{freedman_2021}.  The \shoes team \citep{anand_2024b} has also published an averaged $F090W$ zero point on the Vega-Vega system for the TRGB method, 
giving M$_{F090W}$ = -4.362 $\pm$ 0.033 (stat) $\pm$ 0.045 [sys] mag. Accounting for the approximately 0.035 mag offset between the Vega-Vega and Vega-Sirius systems for $F090W$, their value is -4.327 mag. 

The full range of all three values spans 0.014 mag, suggesting excellent convergence toward an $F090W$ zero point. Furthermore, because the \citet{newman_2024b} $F090W$ value was based on assuming an ingoing $F814W$ zero point equal to -4.05 mag, the remarkable consistency between all three measurements means the \citeauthor{newman_2024b}, \cchp, and \shoes all agree on the TRGB’s $F814W$ zero point being equal to -4.05 mag. These comparisons provide a strong test of the accuracy of the zero point calibration (adopted from F21) that underlies the \hst TRGB distances used in our primary \ho analysis  based on 24 calibrator \sne.\footnote{ Note that if we adopt an $I$-band TRGB zero-point of -3.95 mag, as given in many \shoes papers, the \citeauthor{newman_2024b} $F090W$ zero-point would be 0.1 mag fainter, becoming -4.22 (=-4.32 + 0.1). This value  deviates significantly from the \shoes $F090W$ zero point (-4.36), at greater than a 2-sigma level.}


\subsection{$JWST$ JAGB Measurements}
\label{sec:jagbjwst}

In brief, as described in detail in L25, the JAGB stars in our target galaxies were selected based on their position in (initially blinded) near-infrared ($F115W - F444W$) or ($F115W-F356W$) \jwst \nircam color-magnitude diagrams. The $F115W$ magnitudes were binned and the luminosity functions  were smoothed using a Gaussian-windowed, Locally Weighted Scatterplot Smoothing (GLOESS) algorithm \citep[e.g.,][]{persson_2004}. The JAGB magnitudes for each galaxy were determined from the mode of the smoothed luminosity function.

The JAGB method is best applied in the outer disks and halos of galaxies where  there are  sufficient numbers of carbon stars to provide a statistically meaningful measurement, but not too far into the disk where systematic effects from crowding, blending, and reddening become an issue, and the luminosity function becomes asymmetric. As demonstrated in L25, the JAGB luminosity exhibits a well-defined mode and a distinct Gaussian form in the lower-reddening, less-crowded regions of the galaxies. The mode is also robust against asymmetries in the underlying distribution, as well as robust against outliers in the tails and/or windowing or clipping. The value of the mode is generally found to be brightest in the inner, high-surface brightness regions of a galaxy where the crowding and reddening effects are greatest. For a clean measurement of the JAGB luminosity function, these high-surface-brightness regions need to be avoided. 

The radial cuts for the  JAGB luminosity functions  were set by seeking convergence in the radial distribution: that is, the radial distance, R,  within the disk where the derivative of the magnitude of the JAGB (d$m_{JAGB}$/dR) stabilized and leveled off to zero (L25). In three cases (\ngc 3972, \ngc 4424 and \ngc 4038), a clear convergence was not found, and these galaxies were not included in further analysis. 

\medskip\medskip\medskip\medskip
\subsubsection{Tests for Systematics in Distances Measured Using the JAGB Method}
\label{sec:jagbtests}

To test how our choice of JAGB statistic affected the final measured distances, we explored how the smoothing parameter affected the final measured mode, since the mode measured from an increasingly smoothed luminosity function will eventually converge toward the mean. We varied the smoothing scale of the JAGB luminosity function using smoothing scales of (0.15, 0.20, 0.25, 0.30, 0.35, 0.40) mag, and then re-measured the mode with the given smoothing scale. We defined the statistical error due to the choice of smoothing parameter as the maximum difference between the fiducial mode (measured with a smoothing parameter of $\sigma_s = 0.25$ mag for all galaxies) and all of the measured modes. The total smoothing parameter error was then defined to be the smoothing parameter error from the \sn host galaxy and the zeropoint from \ngc 4258 added in quadrature. This error was larger than 0.05 mag for all galaxies. In \citet{lee_2024c}, we demonstrated that using the mean/median instead of the mode as the chosen JAGB statistic resulted in distance moduli that were measured to be 0.03 mag brighter on average. Thus, this systematic offset was fully encapsulated within the minimum adopted smoothing parameter uncertainty of 0.05 mag. Differences between the mean and the modal values of the JAGB populations were also explored in the study of \citet{madore_freedman_lee_2022}; see their Figure 1.

The effect of a variety of other terms potentially affecting the JAGB distance determination method have been discussed previously in detail in  \citet{freedman_madore_2020}.
These include the effects of star formation history, C-to-M AGB ratio variations, metallicity variations, mass loss, JAGB star variability, foreground/background contamination and lastly, the effect of moving the color-selection window.

For our current study, we adopted a statistical error due to the fluctuations about the final converged JAGB magnitude past the adopted outer disk radial cut. This uncertainty was derived from the dispersion about all measured $m_{JAGB}$ outside of the radial cut, divided by the square root of the number of bins. The presence of a spiral arm in the convergence plots sometimes caused additional noise. As a result,  $m_{JAGB}$  converged at a radial distance with too few JAGB stars or dm/dr not equal to zero in four galaxies: M101, \ngc 2442, \ngc 4258, and \ngc 4639. We masked these spiral arms, and then re-calculated the convergence plots, where $m_{JAGB}$ then successfully converged. We emphasize we performed this procedure during the blinded stage of our analysis. We only masked the spiral arms so that $m_{JAGB}$ successfully converged, independent of knowledge of the final measured distance. However, leaving the spiral arms unmasked while using the newly adopted radial cuts yielded almost a negligible change in \ho  of 0.3\% (larger).

\section{Comparison of the $JWST$ TRGB and JAGB Distances}
\label{sec:comparison}

In Table \ref{tab:distances}, we list the supernovae, host galaxy name, individual distances ($\mu_i$) plus their uncertainties ($\sigma_i$) and  the weighted averages ($\bar{\mu}$) and standard errors $\bar{\sigma}$) for both of the methods. Detailed descriptions of the measurement of these distances and their errors are provided in  H25 (Tables 4 and 5) and L25 (Tables 3 and 5). For the averaging, individual weights are given by  $$w_i = \frac{1}{\sigma_i^2}$$ 
where the weighted mean is 

$$
\bar{\mu _i} = \frac{\sum_{i} w_i \mu_i}{\sum_{i} w_i} 
$$

\noindent
and the weighted standard deviation is 
$$
\sigma_w = \sqrt{\frac{\sum_{i} w_i \cdot (\mu_i - \bar{ \mu })^2}{\sum_{i} w_i}}
$$

 \noindent

For each \sn-host distance average, we incorporated the respective zero-point measurement uncertainty for \ngc 4258: 0.025 mag for the TRGB and 0.041 mag for the JAGB. These errors are correlated  within each method’s distance scale, but uncorrelated  between methods. Consequently, the combined uncertainty of  0.021 mag, calculated as   
 $\frac{1}\sigma_{\text{combined}^2} = \frac{1}{0.041^2} + \frac{1}{0.025^2}$.
 This uncertainty is correlated across all averaged TRGB and JAGB distances. Therefore, we removed it in quadrature from the $\bar{\sigma}$ values in Table \ref{tab:distances}, and  added it back into the final \ho error budget  as a systematic uncertainty.

 Here we compare the individual \jwst JAGB and TRGB measurements. We note that for the galaxy \ngc 5643, there are two \sne that have been observed, SN 2013aa and SN 2017cbv. For the TRGB calibration, there are 11 \sn calibrators that have been observed as
part of Pantheon+ and the \csp.
For the JAGB calibration  three of these are of significantly lower weight: as described in L25, for \ngc 3972, \ngc 4038, and \ngc 4424, no convergence in the radial measurement of the luminosity function was found.

\begin{deluxetable}{llcccccc}
\tablecaption{\jwst Galaxy Distance Moduli   }\label{tab:distances}
\tablehead{\colhead{SN} & \colhead{Galaxy} & \colhead{$\mu_{TRGB}$} & \colhead{$\sigma_T$} & \colhead{$\mu_{JAGB}$} & \colhead{$\sigma_J$} & \colhead{$\bar{\mu}$} & \colhead{$\bar{\sigma}$}}
\startdata
2011fe & M101 & 29.151 & 0.042 & 29.208 & 0.045 & 29.18 & 0.03 \\
2012fr & N1365 & 31.366 & 0.069 & 31.384 & 0.039 & 31.38 & 0.03 \\
2015F & N2442 & 31.646 & 0.097 & 31.605 & 0.044 & 31.61 & 0.03 \\
2011by & N3972 & 31.747 & 0.068 & ... & ... & 31.75 & 0.07 \\
2007sr & N4038 & 31.645 & 0.078 & ... & ... & 31.65 & 0.08 \\
2012cg & N4424 & 30.926 & 0.030 & ... & ... & 30.93 & 0.03 \\
1981B & N4536 & 30.923 & 0.052 & 30.971 & 0.034 & 30.95 & 0.03 \\
1990N & N4639 & 31.774 & 0.073 & 31.733 & 0.039 & 31.74 & 0.03 \\
2013aa & N5643 & 30.643 & 0.071 & 30.582 & 0.038 & 30.60 & 0.03 \\
2017cbv & N5643 & 30.643 & 0.071 & 30.582 & 0.038 & 30.60 & 0.03 \\
2013dy & N7250 & 31.629 & 0.047 & 31.592 & 0.043 & 31.61 & 0.03
\enddata
\end{deluxetable}

\begin{figure}
\centering
\includegraphics[width=10.0cm, angle=-0]{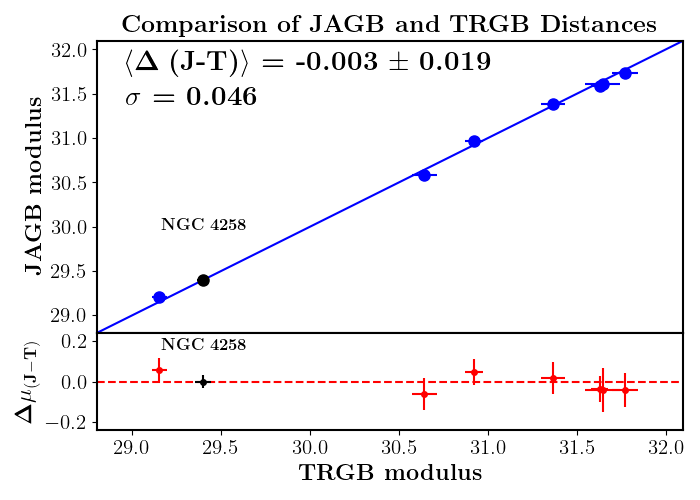}
\caption{ Comparison of JAGB and TRGB distance moduli (blue dots) measured with \jwst \nircam, calibrated with the same anchor galaxy, \ngc 4258. The blue line represents a unit slope. Immediately below, the residuals from the unit slope are shown in red. Two galaxies (\ngc 2442 and \ngc 7250) are close in distance with $\mu \sim 31.6$, and are hard to distinguish (see Table \ref{tab:distances}). }
\label{fig:comparedist}
\end{figure}

An immediate result from our comparison is that the TRGB and JAGB distances are in superb agreement.  This is demonstrated in Figure \ref{fig:comparedist} that compares the eight overlapping JAGB and TRGB distance moduli in our sample (see Table \ref{tab:distances}). The weighted (unweighted) mean difference between the JAGB minus TRGB distance moduli is -0.003 $\pm$ 0.019 (-0.008 $\pm$ 0.018) mag or $<$1\%. The $rms$ scatter about the unit slope is 0.046 mag (2.1\%). In the lower subpanel the residuals from the unit slope line are shown. The agreement for the two independent methods (the JAGB and TRGB distances) is encouraging. 

\medskip\medskip\medskip\medskip\medskip\medskip\medskip\medskip\medskip

\section{Combining Previously-Published Observations of TRGB galaxies: I. HST Data}
\label{sec:trgbprevcompare}

In this section, we describe, compare and combine the TRGB distances from F19 with our new \jwst distances. This doubles our TRGB calibration sample from 10 to 20 galaxies, hosting a total of 24 \sne,  increasing the statistical precision of our TRGB \ho determination. F19 utilized \hstacs F814W ($I$-band) observations to measure 18 \sn calibrators anchored to the Large Magellanic Cloud (LMC). F21 expanded the F19 sample with two new galaxies from \citet{hoyt_2021a},  excluding SN 2007on, underluminous by 3$\sigma$ relative to the calibrator sample, resulting in 19 \sn calibrators.  
Furthermore, F21 incorporated the Milky Way, SMC and \ngc 4258 as additional geometric anchors to the calibration, increasing the total to four. Notably, both F19 and F21 share a common TRGB absolute magnitude zero point, M$_{814W}$= -4.049 mag. Importantly, the four-anchor calibration agrees with the  \ngc 4258 zero point to within 0.001 mag.

Later in \S \ref{sec:trgbcompare}, we compare the TRGB distances for six galaxies in common to the I-band \hst study and our current \jwst $J$-band study. Here we anticipate the excellent agreement at the 0.02~mag (1\%) level. The absence of a significant systematic difference between the F19+F21 \hst and our current H25 \jwst calibrations of the TRGB allows us to combine our \hst and \jwst samples.

H25 further compared our $I$-band TRGB distances with those of \citet{anand_2022} (the Extragalactic Distance Database or EDD), also finding good agreement with a weighted average offset difference of only 0.017 mag, albeit with larger scatter. The \citeauthor{anand_2022} study is based on the same archival \hst data, but the observations have been reduced independently using different software packages. This independent external check provides additional justification for combining our \hst and \jwst TRGB samples.

Hence, both of our \hst and \jwst distances are consistently tied to a common zero point set by the geometric parallax distance to \ngc 4258, which is also in agreement with the more extensive calibration of the \hst distances based on four geometric anchors. 
We list the total sample of CCHP TRGB distances in Table \ref{tab:cchptrgbtot}.  In the cases where more than one TRGB measurement was made per galaxy, we took the inverse weighted average and weighted standard deviation (as described in \S\ref{sec:comparison}) for our \ho analysis.

\begin{deluxetable}{llcccclccc}
\tablecaption{CCHP HST + JWST TRGB and R22 Cepheid Distances} \label{tab:cchptrgbtot}
\tablehead{\colhead{SN} & \colhead{Galaxy} & \colhead{$\mu_{TRGB}^{JWST}$} & \colhead{$\sigma_{TRGB}^{JWST}$} & \colhead{$\mu_{TRGB}^{F19,F21}$} & \colhead{$\sigma_{TRGB}^{F19,F21}$} & \colhead{$\mu_{TRGB}^{CCHP}$} & \colhead{$\sigma_{TRGB}^{CCHP}
$} & \colhead{$\mu_{CEPH}^{R22}$} & \colhead{$\sigma_{CEPH}^{R22}$}}
\startdata
2011fe & M101 & 29.151 & 0.042 & 29.078 & 0.04 & 29.113 & 0.029 & 29.194 & 0.039 \\
2002fk & N1309 & ... & ... & 32.499 & 0.07 & 32.499 & 0.070 & 32.546 & 0.060 \\
2006dd & N1316 & ... & ... & 31.460 & 0.04 & 31.460 & 0.040 & ... & ... \\
1980N & N1316 & ... & ... & 31.460 & 0.04 & 31.460 & 0.040 & ... & ... \\
1981D & N1316 & ... & ... & 31.460 & 0.04 & 31.460 & 0.040 & ... & ... \\
2012fr & N1365 & 31.366 & 0.069 & 31.360 & 0.05 & 31.362 & 0.040 & 31.379 & 0.057 \\
2011iv & N1404 & ... & ... & 31.360 & 0.06 & 31.360 & 0.060 & ... & ... \\
2007on & N1404 & ... & ... & 31.360 & 0.06 & 31.360 & 0.060 & ... & ... \\
2001el & N1448 & ... & ... & 31.320 & 0.06 & 31.321\tablenotemark{a} & 0.038 & 31.290 & 0.037 \\
2021pit & N1448 & ... & ... & ... & ... & 31.321\tablenotemark{a} & 0.038 & 31.290 & 0.037 \\
2015F & N2442 & 31.646 & 0.097 & ... & ... & 31.646 & 0.097 & 31.457 & 0.065 \\
1995al & N3021 & ... & ... & 32.221 & 0.05 & 32.221 & 0.050 & 32.475 & 0.160 \\
1998bu & N3368 & ... & ... & 30.313 & 0.04 & 30.313 & 0.040 & ... & ... \\
1994ae & N3370 & ... & ... & 32.273 & 0.05 & 32.273 & 0.050 & 32.123 & 0.052 \\
1989B & N3627 & ... & ... & 30.221 & 0.04 & 30.221 & 0.040 & ... & ... \\
2011by & N3972 & 31.747 & 0.068 & ... & ... & 31.747 & 0.068 & 31.644 & 0.090 \\
2007sr & N4038 & 31.645 & 0.078 & 31.681 & 0.05 & 31.671 & 0.042 & 31.615 & 0.117 \\
2012cg & N4424 & 30.926 & 0.030 & 31.000 & 0.06 & 30.941 & 0.027 & 30.856 & 0.130 \\
1994D & N4526 & ... & ... & 30.998 & 0.07 & 30.998 & 0.070 & ... & ... \\
1981B & N4536 & 30.923 & 0.052 & 30.964 & 0.05 & 30.944 & 0.036 & 30.838 & 0.051 \\
1990N & N4639 & 31.774 & 0.073 & ... & ... & 31.774 & 0.073 & 31.818 & 0.085 \\
2007af & N5584 & ... & ... & 31.822 & 0.10 & 31.845\tablenotemark{a} & 0.047 & 31.775 & 0.053 \\
2013aa & N5643 & 30.643 & 0.071 & 30.475 & 0.08 & 30.583\tablenotemark{a} & 0.039 & 30.570 & 0.050 \\
2017cbv & N5643 & 30.643 & 0.071 & 30.475 & 0.08 & 30.583\tablenotemark{a} & 0.039 & 30.570 & 0.050 \\
2013dy & N7250 & 31.629 & 0.047 & ... & ... & 31.629 & 0.047 & 31.628 & 0.126
\enddata
\tablenotetext{a}{TRGB distances were obtained for four galaxies (\ngc 1448, \ngc 1559, \ngc 5584 and \ngc 5643) from F090W archival data discussed in \S \ref{sec:newTRGB} (not shown as columns).  }
\end{deluxetable}

\medskip\medskip\medskip\medskip\medskip\medskip\medskip\medskip\medskip
\subsection{Comparison of the Extended TRGB Sample with Riess et al. (2022)}
\label{sec:compareTRGBwithR22}

\begin{figure}
\centering
\includegraphics[width=8.0cm, angle=-0]{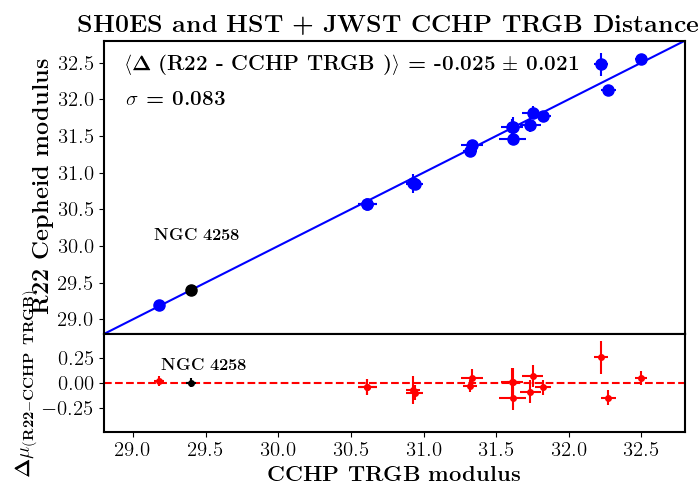}
\caption{A comparison of the distance moduli measured using Cepheids (R22) with the distance moduli measured using the TRGB (from F19, F21 and this paper). The R22 Cepheid distances now agree well with the distances measured using the TRGB, as published by F19). The offset (in the sense of Cepheids minus TRGB) amounts to -0.025 $\pm$ 0.021 mag, which corresponds to a mean difference in distance of 1.2\%. 
\label{fig:compare_R22_TRGBtot}}
\end{figure}

Figure \ref{fig:compare_R22_TRGBtot}  compares our combined    \hst plus \jwst TRGB distances with the R22 Cepheid distances for 17 common galaxies. The agreement is excellent, with a weighted mean difference of only 0.025 $\pm$ 0.021 mag, corresponding to a  1.2\% difference in distance. The scatter is 0.08~mag, or 4\% in distance.  
The offset, at 1.2-$\sigma$, is consistent with zero. Both the \cchp TRGB and R22 Cepheid distance moduli for these galaxies are listed in Table \ref{tab:cchptrgbtot}.

This agreement in nearby distances is a noteworthy development. Previously, F19 reported a weighted average distance modulus difference (TRGB minus Cepheid) of +0.059~mag between the TRGB and \citet{riess_2019} distances. F19 noted also that the scatter in the local Hubble diagram was larger for the Cepheid distances, indicating that the discrepancy was not equally distributed between the two methods, suggesting that the TRGB distances had higher precision than the Cepheid distances. This 3\% distance difference partially accounted for the  \ho discrepancy at the time (\ho = 73.24 $\pm$ 1.74 and \ho = 69.8 $\pm$ 0.8 (stat) $\pm$ 1.7 (sys), for Cepheids and the TRGB, respectively). As outlined in F19, the differences in \ho stemmed from: 1) the differences in distance moduli between the TRGB and Cepheid distances, 2) differing \sne calibrators, and 3) varying weights assigned to  \sn calibrators, arising from different uncertainties in distance and \sn measurements.

The first reason, the 0.06 mag discrepancy between the F19 TRGB and the R16 Cepheid distance scales, is now (partially) resolved as a source of the disagreement. As we have seen, the re-analysis by R22 of the Cepheid distances now results in distances to the \sne that agree to within 0.025 mag of those from both the earlier F19 study and from our new \jwst \cchp data (L25, H25).

\section{Calibration of \sne and the Hubble Constant}
\label{sec:h0}

None of the three methods  in our \jwst program (Cepheids, TRGB or JAGB stars) are  bright enough to  determine distances out into the smooth cosmic Hubble flow at a level of 1\% accuracy in \ho. Nearby galaxies and clusters induce motions  or peculiar velocities, scattering above and below the Hubble expansion velocities, adding noise and potential bias to the determination of \ho.  In the past couple of decades, \sne have surfaced as the preferred secondary distance indicator given their high intrinsic brightness and their small observed dispersion in the Hubble diagram ($\pm$0.1-0.15 mag, e.g., \citep{burns_2018, scolnic_2022,brout_2022}). In this paper, we apply our independent  TRGB and JAGB star distances to two samples of distant supernovae: 1) the (\csp) \citep{uddin_2024} and 2) the Pantheon+ sample \citep{scolnic_2022}, as described below. 

\subsection{The Carnegie Supernova Program (CSP)}
\label{sec:CSP}

The \csp \citep{contreras_2010, folatelli_2010, krisciunas_2017} was initiated 20 years ago as a program to provide multi-wavelength follow-up observations of previously discovered supernovae.  The data were intended for applications to cosmology, as well as for studying the physical properties of the supernovae themselves\footnote{The \csp data are available at \url{http://csp.obs.carnegiescience.edu/data.}}. The aim was to obtain homogeneous, intensive, high-cadence, multi-band $uBVgriYJH$ observations  \citep{contreras_2010}.  Careful attention was given to photometric precision and systematics: the program utilized a fixed set of instruments, photometric standard stars, and instrumental reduction procedures, catching most of the supernovae well before maximum, and with high signal-to-noise, avoiding many of the challenges otherwise faced in minimizing systematic differences across multiple data sets/instruments/etc. \citep{krisciunas_2017}. Optical spectra were also obtained with high cadence \citep{folatelli_2013,morrell_2024}. 
The bulk of the observations were carried out at Las Campanas Observatory using the 1-m Swope and 2.5-m du Pont telescopes. The first part of the \csp (\cspi) was carried out from 2004-2009, with a second phase (\cspii) from 2011-2015, optimized for the near-infrared \citep{phillips_2019, hsiao_2019}. Light curves were generated using the analysis package SNooPy \citep{burns_2018}. This program determines the peak magnitude in the light curves for each of the $uBVgriYJH$ filters, the times of those maxima, (B-V) colors, and a color-stretch parameter s$_{BV}$, as described by \citet{burns_2014}. The color-stretch parameter  incorporates color in the relation between the luminosity and decline rate of a \sn, allowing also for the redder colors of fast decliners. 
Finally, it computes the covariance matrix involving all of the parameters.

Previous applications of the \cspi survey to cosmology include \cite{burns_2018}, who used the \cspi \sne sample, calibrated using the Cepheid distances of \cite{riess_2016}, to obtain a value of \ho = 73.2 $\pm$ 2.3 \hounits (for H-band data); and a value of \ho = 72.7 $\pm$ 2.1 \hounits (for B-band data). Calibrating the \cspi sample with the TRGB \citep{freedman_2019}, updated in \citep{freedman_2020, freedman_2021} found a lower value of \ho = 69.8 $\pm $ 0.6 (stat) $\pm$ 1.6 (sys) \hounits.  \cite{uddin_2024} have updated the \csp analysis, including the more recent \cspii \sne data, which, once included with \cspi, triples the sample size over \cspi. They undertook a calibration based on Cepheids, the TRGB and Surface Brightness Fluctuations, the latter being a secondary distance indicator calibrated first by Cepheids. Using B-band light-curve fits, they find \ho = 73.38 $\pm$ 0.73 \hounits based on the Cepheid calibration of \cite{riess_2022}. For the TRGB calibration, they find \ho = 69.88 $\pm$ 0.76 \hounits, based on the TRGB distances of  \cite{freedman_2019}. Both of these later results are in excellent agreement with the original studies. 

\subsection{JWST CSP calibration}
\label{sec:jwstcsp}

The \sne data used in this work are those described in the recent analysis by \cite{uddin_2024}. There are over 300 \sne in this 
sample, including both \cspi and \cspii at $BVri$ wavelengths, and more than 200 \sne observed at $JH$. In analyzing these data, we follow the methodology described previously by \cite{burns_2018, freedman_2019, uddin_2024}.
As in \citeauthor{burns_2018}, the corrected magnitudes for an individual filter (e.g., the B filter below) are given by:

\begin{equation}
\label{eq:m_Tripp}
B_{corr}  =  P^0 - P^1(s_{BV}-1) - P^{2}(s_{BV}-1)^2 - \beta(B-V) - 
       \alpha_M\log_{10}({M_*-M_0\over M_\odot})
\end{equation}

\noindent
where in this case $P^0$ = $B$, the apparent peak (K-corrected) $B$ magnitude, $P^1$ is the linear coefficient, and $P^2$ is a quadratic coefficient in ($s_{BV}-1$) where $s_{BV}$ is the stretch parameter described in \S\ref{sec:CSP}; $\beta$ is the slope of the color correction; $V$ is the apparent peak magnitude at $V$, K-corrected; $\alpha_M$ is the slope of the correlation between peak luminosity and host stellar mass $M_*$; and M$_0$ is the median value of the host stellar mass. Host stellar masses are derived as described in \cite{uddin_2024}. Following \citeauthor{uddin_2024}, the sample is split at the median mass so that equal weights are given above and below the median mass (although the results are not significantly affected by the choice of split point).

The apparent magnitudes at maximum are computed by fitting the light curves with SNooPy, providing the time of maximum, the light-curve shape $s_{BV}$, and the magnitude at maximum luminosity for each filter. These quantities are then provided as inputs to a Markov Chain Monte Carlo (MCMC) sampler that simultaneously solves for all the correction factors: $P^1$, $P^2$, $\alpha_M$, and $\beta$  \citep[for full details, see][]{burns_2018, uddin_2024}. The MCMC sampler then provides the corrected magnitudes, as well as a full covariance matrix, which is used when determining \ho and its error.

The distance modulus is defined as in \citet{burns_2018} for a flat cosmology:
\begin{equation}
\label{eq:mu}
\mu(z,H_0,q_0) =  {\rm 5~log_{10}} \lbrace  {{(1+z_{hel})} cz \over{{(1+z)}H_0}} (1 + {1-q_0\over{2}}z )\rbrace  + 25
\end{equation} \linebreak
\noindent
where  $q_0$ is the deceleration parameter, and z is the redshift in the CMB frame.

As described in \cite{uddin_2024}, three error terms are included in the analysis. The first, $\sigma_i$ is the sum of the individual errors for the observed quantities, in addition to the covariance between the peak magnitude, and both the color and the color-stretch parameter. The second, $\sigma_{int}$ is the intrinsic random scatter that allows for variations beyond either $\sigma_i$ or those due to peculiar velocities, $\sigma_{pec}$. The third term is the error due to the uncertainty in the distance measurements resulting from galaxy peculiar velocities, and scales with redshift. \cite{burns_2018} define
$\sigma_{pec} =  2.17 \times  V_{pec} / cz_{cmb}$  where V$_{pec}$  represents the average peculiar velocity of the \sn sample.  We set the average peculiar velocity V$_{pec}$ = 240 km s$^{-1}$, a value consistent with that determined by \citet{brout_2022, scolnic_2015, burns_2018}. In practice, this made an insignificant difference in the determination of \ho. Adopting different values, taken in steps of 50 km s$^{-1}$ in the range of 150 $< V_{pec} < 350$ km s$^{-1}$ resulted in differences in \ho at a level $<<$1\%.

There are some differences in the \cspi and \cspii samples previously noted by \citet{uddin_2024}. 
The combined CSP-I plus CSP-II sample results in an increase by almost a factor of three relative to the CSP-I sample alone. The scatter in the \cspii data is larger than for \cspi, which as noted by  \citeauthor{uddin_2024}, results from a doubling of the number of low-redshift \sne in the combined \csp sample. The average peculiar velocity (V$_{pec}$) for the \cspii sample was also larger than found for \cspi.  
In this paper, we consider only redshifts  of z $>$ 0.01 \citep[see e.g.,][]{brout_2022, chen_2024} to limit effects of peculiar velocities for nearby objects. We explored redshift cuts ranging from  0.0065 $<$   z $<$  0.03, and found differences with a full range of only 1\% in \ho. These differences were stochastic, and showed no trend with increasing redshift cuts. Finally, the $\beta$ parameter is also steeper in the combined sample than the \cspi sample. One possibility is that the combined sample could contain more star-forming galaxies with larger amounts of dust. These issues highlight the fact that in the future, in an era where a value of \ho at a level of 1\% accuracy is the goal, large {\it and homogeneous} samples of \sne become increasingly important.

\subsection{The Hubble Diagram}
\label{sec:hubdiag}

\begin{figure*} 
 \centering
\includegraphics[width=1.0\textwidth]{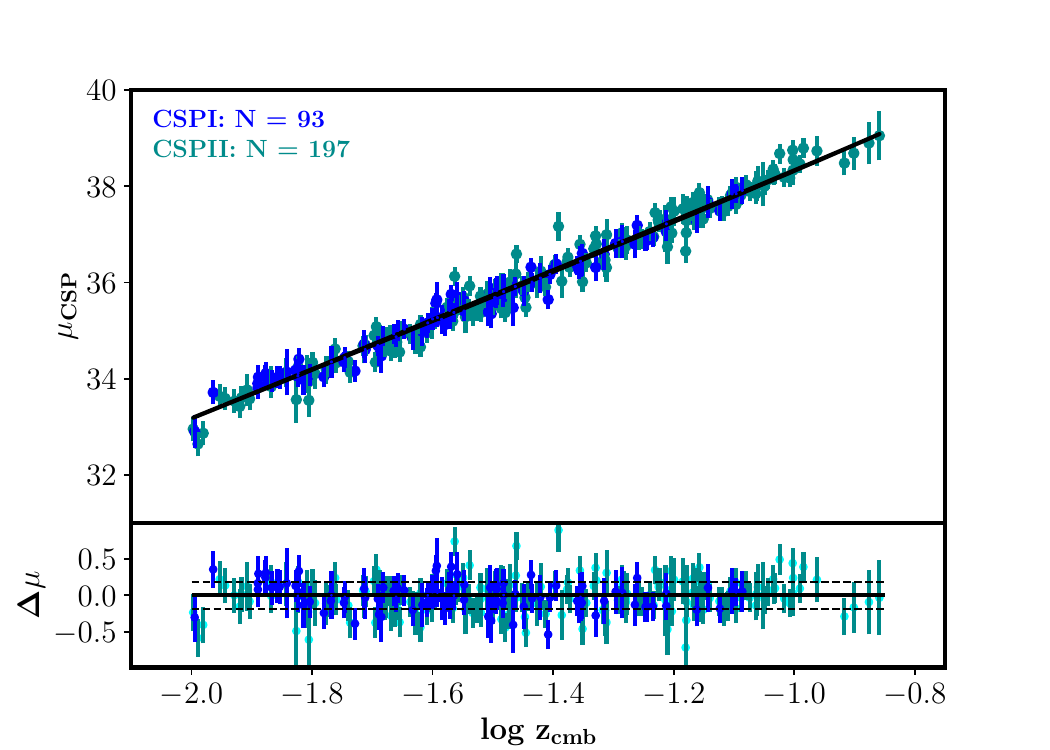}
 \caption{A Hubble diagram for 287 \sne observed as part of the \csp (blue filled circles (\cspi) and dark cyan filled circles (\cspii). 
 A slope = 5 line is plotted.  The lower panel shows residuals about the slope = 5 line. 
}
\label{fig:CSPHubble}
\end{figure*}

In Figure \ref{fig:CSPHubble}  we show the Hubble diagram for the 
\csp sample of 287 SNe Ia (blue filled circles (\cspi) and dark cyan filled circles (\cspii) where z$>$0.01, while  excluding super-Chandra (IA-SC) and IaX subtypes. In addition, two additional 3$\sigma$ outliers (SN2014D and SN2013hh) were excluded from the analysis.
The calibrating galaxies provide distances only; their velocities are not used in the calibration of \ho. (That is, these galaxies would appear in Figure \ref{fig:CSPHubble} to fall on the trend line with zero dispersion. The calibrating \sne constrain only the intercept.)  Residuals from the fit in distance modulus  are shown in the lower panel.

\subsection{Markov Chain Monte Carlo Analysis}
\label{sec:mcmc}

 We explore the use of two  Python packages, {\ttfamily emcee} \citep{foreman-mackey_2013} and {\ttfamily pymc3} \citep{pymc_2023}, and find that they give consistent results. 
 Initially, using {\ttfamily emcee}, a set of broad, uniform priors were adopted for 8 parameters (P$^0$, P$^1$, P$^2$, $\beta$, $\alpha$, V$_{pec}$, $\sigma$, and \ho) and 100 walkers were used. The burn-in time was set to be 5 times the autocorrelation time, $\tau$, generally 500 steps, and the acceptance fraction was generally $\sim$0.45-0.49, within the range of 0.2-0.5 recommended by \citet{gelman_1996}. The number of chains after burn-in was taken to be greater than 100$\tau$, and convergence was visually checked in walker trajectory plots. The adopted parameter fits are given by the marginalized distributions and uncertainties quoted are 16th, 50th and 84th percentiles. In the case of the second analysis using  {\ttfamily pymc}, 30,000 steps were implemented, and the burn-in time was set to 3,000. 
 {\ttfamily pymc} uses ``forward sampling" using the derivatives of the probability hyper-surface. Consistent results were obtained in both cases. 

 We note that the sample of ten galaxies for which we have \jwst distances (and 11 \sne)  is currently small, resulting in a larger statistical uncertainty relative to longer-running \hst-based studies ({\it e.g.}, F19, R22). The $rms$ scatter in the peak \sn magnitudes for this nearby sample is smaller than for  the distant \csp sample. Allowing only for a single parameter to account for the scatter, $\sigma_{int}$, leads to an anomalously small dispersion in the P0 posterior, and hence, the \ho posterior. We thus included an additional 8th parameter ($\sigma_{cal}$), allowing for this difference in the scatter between the calibrating and distant sample, and thereby allowing for a larger uncertainty  in \ho. The scatter in P$^0$ then increased by a factor of five, and the resulting value of \ho increased by 1\%, with a two-fold increase in its uncertainty.

Table \ref{tab:mcmc} presents the parameter fits derived from our {\ttfamily pymc} analysis, encompassing results for \jwst TRGB, JAGB, and combined TRGB+JAGB distances, as well as the full \hst+\jwst TRGB sample. Using B-band \csp \sn data and the new \jwst data, we obtained the following peak marginalized \ho distributions: 68.81 $\pm$ 1.79 \hounits for the TRGB and 67.80 $\pm$ 1.49 \hounits for JAGB (statistical errors). These values are based on 11 \sn calibrators for TRGB and 8 for the JAGB. In an earlier version of this paper, the analysis did not include SN 2017cbv, which was observed after the \cspii program had been completed. SN 2017cbv was, however, observed with exactly the same instrumentation and optical bandpasses as for \cspii \citep[see][]{burns_2020}.

The statistical uncertainty for the JAGB method, as listed in Table \ref{tab:mcmc}, is smaller than that for TRGB. This is due to the blinded JAGB sample excluding three supernovae (SN 2011by, SN 2007sr, and SN 2012cg) that are included in the TRGB sample. These omissions, resulting from radial measurements not converging for these galaxies, led to an artificially small dispersion in the JAGB-\sn calibration. To avoid underestimating the JAGB uncertainty, we increased it by a factor of 2.45, the ratio of the calibration dispersions between TRGB and JAGB.

A {\ttfamily pymc} analysis of the weighted mean of \jwst TRGB and JAGB distances (Table \ref{tab:distances}, column 7) yields an \ho value of 69.02 $\pm$ 1.66 \hounits (statistical error). The differences in \ho between TRGB, JAGB, and the combined sample arise from their respective uncertainties and weightings. Notably, the averaged distance moduli for \ngc 5643 and \ngc 2442 are significantly lower in the combined sample than for the TRGB alone, contributing to a slight (0.3\%) increase in \ho.

\begin{figure}[ht!]
\plotone{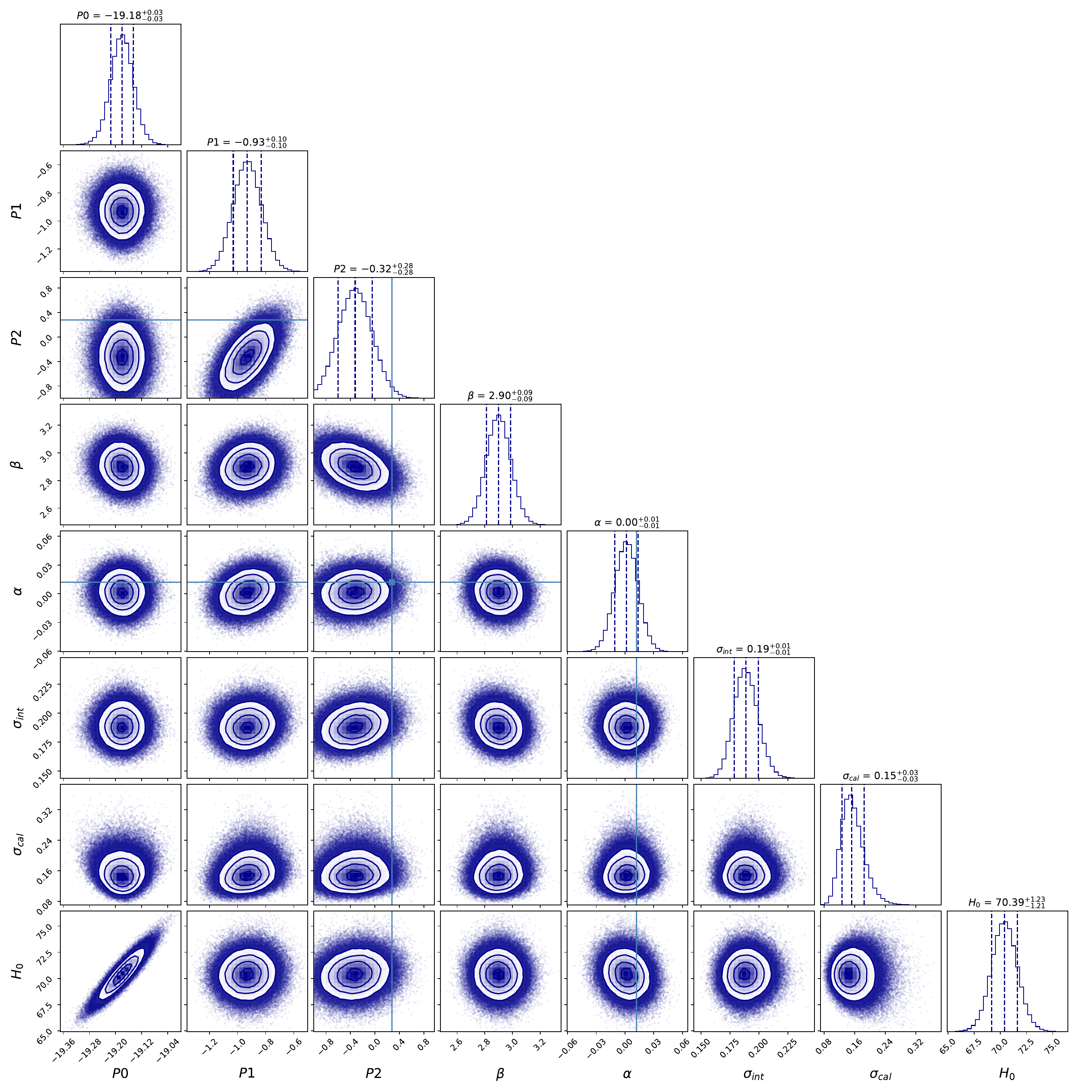}  
\caption{Corner plot with  two-dimensional projections and one-dimensional marginalizations of the posterior probability distributions of the {\it \bf pymc} fitting parameters. This plot shows an example of the \hst and \jwst calibration of the B band for  24 \sne based on} the TRGB method
as given in Table \ref{tab:distances}, and  the analysis described in Section \ref{sec:jwstcsp}. Parameter values are given in Table \ref{tab:mcmc}. The plots correspond to a 30,000 step run with the {\it \bf pymc} sampler. 
\label{fig:corner}
\end{figure}

Expanding the analysis to the larger combined \hst + \jwst TRGB sample of 24 \sne provides our best current estimate, yielding \ho =  70.39  $\pm$ 1.22 \hounits (statistical error).  We show an example corner plot from  {\ttfamily pymc} for the 24 TRGB calibrators in Figure \ref{fig:corner}. 

The \ho values obtained from the TRGB sample (68.81 $\pm$  1.79) and the full HST+JWST sample (70.39 $\pm$ 1.22) are statistically consistent. Eleven of the 24 supernovae are common to both samples, and accounting for uncorrelated errors in the independent measurements results in a difference at the 0.8 $\sigma$ level ((70.39 - 68.81) / $\sqrt{1.8^2 + 0.7^2}$).

A discussion of the systematic uncertainties is deferred to a later section of the paper.
We  keep the statistical and systematic errors separate,  so that they can be easily identified.

\begin{deluxetable*}{lcccc}
\tablecaption{{\ttfamily pymc} MCMC Parameter Output}
\label{tab:mcmc}
\tablehead{
\colhead{Parameter} &
\colhead{TRGB$_{tot}$} &
\colhead{TRGB$_{JWST}$} &
\colhead{JAGB$_{JWST}$}  &
\colhead{TRGB+JAGB} 
}
\startdata
No. of calibrators   &   24   & 11  & 8 & 11 \\
P$^0$ (mag)  &   -19.18 $\pm$ 0.04  & -19.23 $\pm$ 0.05 & -19.26 $\pm$ 0.05 & -19.22 $\pm$ 0.05\\
P$^1$ (mag)  & -0.93 $\pm$ 0.10 &  -0.91 $\pm$ 0.10 & -0.91 $\pm$ 0.10 & -0.91 $\pm$ 0.10\\
P$^2$ (mag)  & -0.32 $\pm$ 0.28 &  -0.32 $\pm$ 0.29  & -0.32 $\pm$ 0.29 & -0.32 $\pm$ 0.29 \\
$\beta$ & 2.90 $\pm$  0.09 & 2.92 $\pm$ 0.09 & 2.92 $\pm$ 0.09 & 2.92 $\pm$ 0.09 \\
$\alpha$ (mag/dex) & 0.00 $\pm$ 0.01 & 0.00 $\pm$ 0.01 & 0.01  $\pm$ 0.01 & 0.01 $\pm$ 0.01 \\
$\sigma_{\rm cal}$ (mag) & 0.15 $\pm$  0.03 &  0.16 $\pm$ 0.05 & 0.11 $\pm$ 0.05 &  0.15 $\pm$ 0.05 \\
$\sigma_{\rm int}$ (mag)  & 0.19 $\pm$ 0.01 & 0.19 $\pm$ 0.01  &  0.19 $\pm$ 0.01 & 0.19 $\pm$ 0.01\\
H$_0$ (\hounits) &  70.39 $\pm$ 1.22 & 68.81 $\pm$ 1.80 & 67.80 $\pm$  2.17\tablenotemark{a} & 69.02 $\pm$ 1.56 
\enddata
\tablenotetext{a}{As described in the text,  the uncertainty for the JAGB method was increased by a factor of 1.45 (given by the ratio of the $\sigma_{cal}$ value for TRGB  to that of the lower $\sigma_{cal}$ value for the JAGB), so that the JAGB method does not unduly contribute more weight simply because of the smaller dispersion in its smaller number of calibrators. }
\end{deluxetable*}

As previously noted,  the total \csp sample includes a significant fraction of low-redshift \sne which, if included in the MCMC analysis,  would require a correction for the local-density-field velocity perturbations. For example, based on predictions from the 2M++ density field of \cite{carrick_2015}, \cite{uddin_2024} corrected the \csp peculiar velocity measurements (for which the average peculiar velocity is $\sim$440 km s$^{-1}$). They found a mean sample velocity correction for the \csp survey of 90 km s$^{-1}$, resulting in a net increase in \ho of 0.55 \hounits in the B band. However, as shown in \citet{brout_2022}, not including supernovae at z $<$ 0.01 avoids the bias due to nearby peculiar velocities. Also as previously noted, our preferred value of \ho is based on a sample of 287 \sne with z$~>~$0.01, eliminating also  super-Chandra (IA-SC) and IaX subtypes, thus requiring no additional correction to \ho.

In Table \ref{tab:hovalues} we summarize the values of \ho obtained using the two methods applied in this paper, based on the \csp \sn data. We show also values that are based on the Pantheon+ \sn data (see discussion in \S\ref{sec:pantheon} below), for which there is good agreement. We present the results from different MCMC analyses ({\ttfamily pymc} and {\ttfamily emcee}), show the results adopting different redshift cuts for the \csp \sne sample, adopting different nearby calibrators (excluding those which show the largest deviations, for example), and for calibrating the $H$-band data for the \csp \sne. We have also examined the difference in \ho when excluding \sne observed as part of the \csp, but not the \shoes analysis; i.e., \sne that occurred in elliptical or S0 galaxies. To within the uncertainties, the various cuts do not lead to significant variations in \ho, and all are consistent with a value of \ho = 70 \hounits. In no case do we find a value of \ho of 73 or 74 \hounits.    We note that the total \csp sample of \sne with $H$-band data is not as large as that for the $B$ band (213 compared to 322), and the values of \ho obtained based on the  $H$-band  are larger than for the $B$ band. \citet{uddin_2024} searched, but found no evidence, for potential calibration errors. There are, however, hints of the existence of a brighter sub-population of \sne in the near-infrared, which warrants further investigation. The uncertainties in these values are discussed in \S\ref{sec:overall_systematics}. 

Our adopted value of \ho, based on the more extensive $B$-band data, and the combined (\hst + \jwst) TRGB sample is listed at the end of the table. We do not apply a peculiar-velocity model correction to this value of \ho. Uncertainties in model-dependent corrections are difficult to quantify, and may be as large, or larger, than the corrections themselves.\footnote{For example, in the extensive fundamental plane study undertaken as part of the DESI Peculiar Velocity Survey, in their Figure 7,  \citet{said_2024} plot several velocity models \citep[e.g.,][]{carrick_2015} in comparison to the residuals from their Hubble diagram, none of which provide a good fit to the data.} \citet{uddin_2024} apply a 0.55 \hounits correction to their value of \ho; here we absorb the uncertainty in the velocity modeling as an additional systematic uncertainty. To be conservative, we take this uncertainty, $\sigma_{SN}$, to be an additional 1\%, which also provides allowance for remaining uncertainty in the photometric calibration for the \sne.

\begin{deluxetable*}{lclc}
\tablecaption{Summary of \ho Values and Statistical Uncertainties  \label{tab:hovalues}} 
\footnotesize 
\tablehead{\colhead{\ho\tablenotemark{a}} & \colhead{ Error\tablenotemark{a}} & \colhead{Description} & \colhead{MCMC code}} 
\startdata
\multicolumn{4}{c}{\bf TRGB} \\ 
\cline{1-4}  
70.39      &       +1.23 -1.21       &       TRGB + \csp B band ; 24 SN calibrators, z$>$0.01 ; 291 distant SN  &    {\ttfamily pymc}  \\
70.39      &       +1.23 -1.22      &       TRGB + \csp B band ; 24 SN calibrators, z$>$0.0065  ;  303 distant SN &    {\ttfamily pymc}  \\
70.48      &       +1.24 -1.22      &       TRGB + \csp B band ; 24 SN calibrators, z$>$0.015  ;  264 distant SN &    {\ttfamily pymc}  \\
70.37      &       +1.26 -1.25       &       TRGB + \csp B band ; 24 SN calibrators, z$>$0.02  ; 235 distant SN &    {\ttfamily pymc}  \\
69.81      &       +1.32 -1.29       &       TRGB + \csp B band ; 24 SN calibrators, z$>$0.025  ; 201 distant SN &    {\ttfamily pymc}  \\
70.11      &       +1.24 -1.23       &       TRGB + \csp B band ; 23 SN calibrators, omit SN2007on, z$>$0.01   &    {\ttfamily pymc}  \\
70.01      &       +1.21 -1.19       &       TRGB + \csp B band ; 23 SN calibrators, omit SN2011by, z$>$0.01   &    {\ttfamily pymc}  \\
70.73      &       +1.21 -1.18       &       TRGB + \csp B band ; 23 SN calibrators, omit SN1981D  z$>$0.01   &    {\ttfamily pymc}  \\
70.45      &       +1.23 -1.19       &       TRGB + \csp B band ; 22 SN calibrators, omit SN1981D and 2007on, z$>$0.01   &    {\ttfamily pymc}  \\
69.96      &       +1.18 -1.16       &       TRGB + \csp B band ; 23 SN calibrators, omit SN2007af, z$>$0.01   &    {\ttfamily pymc}  \\
70.50      &       +1.54 -1.49       &       TRGB + \csp B band ; 16 SN calibrators, omit SNe not observed by R22\tablenotemark{b}, z$>$0.01   &    {\ttfamily pymc}  \\
68.81      &       +1.80 -1.77       &       TRGB + \csp B band ; 11 SN calibrators, \jwst distances only, z$>$0.01   &    {\ttfamily pymc}  \\
70.43       &       +1.22 -1.30       &       TRGB + \csp  B band ; 24 SN calibrators, z$>$0.01   &    {\ttfamily emcee}  \\
\hline
\multicolumn{4}{c}{\bf JAGB} \\ 
\cline{1-4}
68.75       &       +1.56 -1.52       &       JAGB + \csp B band ; 11 SN calibrators, z$>$0.01    &    {\ttfamily pymc}    \\
67.80       &       +1.50 -1.48       &       JAGB + \csp B band ; 8 SN calibrators\tablenotemark{c}, z$>$0.01    &    {\ttfamily pymc}     \\
\hline
\multicolumn{4}{c}{\bf TRGB + JAGB Distances Averaged  (z$>$0.01, \csp B-band)} \\ 
\cline{1-4}
69.02     &       1.69 -1.63     &       Averaged (TRGB, JAGB) distances ; 11  SN calibrators   &   {\ttfamily pymc}     \\
67.60      &       1.58 -1.51     &       Averaged (TRGB, JAGB) distances ; 8  SN calibrators   &   {\ttfamily pymc}     \\
\hline
\multicolumn{4}{c}{\bf Pantheon \sne sample  } \\ 
\cline{1-4}
69.29      &       ...      &       TRGB + Pantheon+ B band  ; 11 SN calibrators  &  \shoes\tablenotemark{d}    \\
68.80      &       ...       &       JAGB + Pantheon+  B band  ; 8 SN calibrators  &   \shoes\tablenotemark{d}     \\
\hline
\multicolumn{4}{c}{\bf Adopted Result} \\ 
\cline{1-4} 
\multicolumn{4}{c}{\bf 70.39              $\pm$1.22 (stat)        $\pm$ 1.33 (sys) $\pm$ 0.70 ($\sigma_{SN}$) }            
\enddata
\tablenotetext{}{(a) Units of \hounits; errors quoted are statistical, except for the combined and adopted result where both statistical and systematic errors are separately tabulated. }
\tablenotetext{}{(b) Excluding SN 2007on, 1981D, 1989B, 1998bu, 1994D, 1980N, 2006dd, 2011iv in galaxies not observed by R22.  }
\tablenotetext{}{(c) Excluding SN 2011by in \ngc 3972, SN SN2007sr in \ngc 4038 and SN2012cg in \ngc 4424.}
\tablenotetext{}{(d) Here the \cchp zero points, based on the peak $B$ magnitudes for the calibrating sample, are applied to the R22 covariance matrix analysis for \ngc 4258. R22 find \ho = 72.51 $\pm$ 1.54 for \ngc 4258, but do not list the statistical and systematic uncertainties separately.  }  
\end{deluxetable*}

In summary, the final \ho value that we adopt for this paper is based on the \hst + \jwst TRGB sample of 24 \sn calibrators,  tying into the $B$-band \csp \sn sample  with \ho = 70.39 $\pm$ 1.22 (stat) $\pm$ 1.33 (sys) $\pm$ 0.70 [$\sigma_{SN}$]. We note the good  agreement of the \ho values using only the \jwst TRGB and JAGB  distances, as well as  their averaged distances, despite the lower statistical power of the \jwst samples alone.

\subsection{Comparison with Pantheon+ }
\label{sec:pantheon}

In the Pantheon+ analysis, \cite{scolnic_2022} collected and cross-calibrated the data for 1550  individual \sne,  superseding earlier Pantheon \citep{scolnic_2018} and Joint Light-Curve \citep{betoule_2014} analyses. The total sample includes \sne in the redshift range 0 $<$ z $<$ 2.3. The analysis aims to standardize the B-band photometry from 18 individual surveys obtained with a wide variety of telescopes and instruments\footnote{ The Pantheon+ catalog is available at \url{https://github.com/PantheonPlusSH0ES/DataRelease.}}. In the \shoes determination of \ho, \citep[e.g.,][]{riess_2009, riess_2022}, a simultaneous fit is undertaken for the Cepheid and \sne data,  minimizing a $\chi^2$ statistic, and most recently providing the covariances. R22 also utilize {{\ttfamily emcee} as a check of their methodology, finding almost exact agreement.  The \shoes Cepheid calibration of the Pantheon+ \sne sample from \cite{scolnic_2022} results in a value of \ho = 73.04 $\pm$ 1.04 \hounits for the 277 \sne with 0.023 $<$ z $<$ 0.15.

 For comparison purposes, to be completely consistent with the \sn analyses undertaken by \citet{scolnic_2022} and R22, we do not re-analyze the Pantheon+ data. We determine \ho by scaling the absolute M$_B$ values relative to those measured by R22 (their 'baseline' fit) where M$_B$ = -19.253 $\pm$ 0.027 mag, allowing for the fact that our calibration is based on \ngc 4258 alone, and not the baseline calibration of the LMC, Milky Way and \ngc 4258, which corresponds to value of \ho = 73.04 $\pm$ 1.01 \hounits. The R22 value of \ho based on \ngc 4258 alone is  72.51 $\pm$  1.54 \hounits. That is,
 
 \begin{equation}
\label{eq:weights}
M_B (N4258)  = M_B(\textrm{R22 baseline) - 5 log} { H_0 (\textrm{R22 baseline)} \over H_0 (N4258) 
} 
= \textrm{-19.269}
\end{equation}

 \noindent
 We note that this value is consistent with that of R22, Table 5, entry \#10, but differs from that quoted by R24b who quotes instead  M$_B$ = -19.28.  Adopting a value of M$_B$ = -19.269 for the Pantheon+ data we find 
 \ho = 69.29 for the TRGB and \ho = 68.80  \hounits for the JAGB. These values are in excellent agreement with those determined based on the \csp 
\sne calibration.

\subsection{Absolute Magnitude M$_B$ Distributions for Pantheon+ SN Ia Calibrators }
\label{sec:compareMBhists}

 
In Figure \ref{fig:SH0EScomparePDFs} we compare the absolute magnitude \sn peak brightness, M$_B$, for the set of \sne contained in the Pantheon+ sample. There are 20 galaxies for which we have 
measured TRGB  distances (hosting 24 \sne), and 7 galaxies with well-measured JAGB distances (hosting 8 \sne). The  weighted mean magnitude and error on the mean for the total sample is $\overline{M_B}$ = -19.316  $\pm$ 0.033 and -19.339 $\pm$ 0.029 mag (with and without SN 2007af, respectively). This single supernova has a disproportionate effect: including SN2007af increases \ho by 1\%; i.e., it moves the weighted average by nearly the entire size of the R22 error bar. These values can be compared to the overlapping sample of R22 \sn discussed above where $\overline{M_B}$ 
= -19.317  $\pm$ 0.033 mag. 
Finally, the  weighted mean magnitudes and error on the mean  are  $\overline{M_B}$ =  -19.363 $\pm$ 0.026~mag (TRGB), and -19.377 $\pm$ 0.023~mag (JAGB), respectively. 

\begin{figure*} 
 \centering
\includegraphics[width=0.5\textwidth]{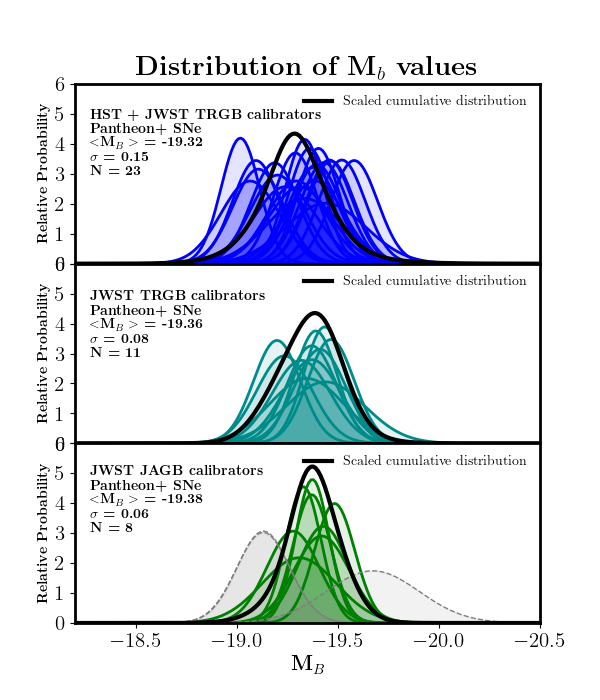}
 \caption{ Probability distribution functions (PDFs) of M$_B$ values for \sne based on the \hst + \jwst TRGB sample and the \jwst-alone  calibrations of the  TRGB and JAGB  from this paper. The \sn data are from Pantheon+, as given in R22. The cumulative distributions are shown in black. The highest-weight (and faintest) supernova in the total TRGB sample is SN 2007af in \ngc 5584. The gray (dashed) curves in the bottom JAGB plot represent the PDFs for SN2007sr,  SN2011by, and SN2007cg, which have peak M$_B$ magnitudes of -19.131, -19.132 and -19.673. They are located  in the galaxies \ngc 4038, \ngc 3972, and \ngc 4424, respectively, for which the peak of the JAGB luminosity function did not converge (see \S\ref{sec:jagbjwst}). The former two \sne have nearly identical values of M$_B$, and indistinguishable PDFs.  }
 \label{fig:SH0EScomparePDFs}
\end{figure*}

We show in Figure \ref{fig:MBhistCSP} a histogram of absolute magnitude (M$_B$) values for the 24 host-\sne calibrating galaxies  in the \cchp sample. The TRGB distances are those from this paper and F19, and the apparent \sn magnitudes (m$_B$) values are those adopted by R22. Shown in yellow are the \jwst data from this paper. The expanded \hst + \jwst TRGB sample of 24 \sne now exhibits a distribution more representative of the larger R22 sample of 42 \sne, consistent with the expected `reversion to the mean' for larger datasets. This increased sample size demonstrates that the mean of the TRGB-calibrated M$_B$ sample has remained stable. {\it The criticism from \citet{riess_2024b} [hereafter (R24)] regarding our earlier, smaller 
\jwst-only sample is no longer applicable, as it is superseded by our current analysis of 24 \sne using \hst and \jwst data.}

\begin{figure*} 
 \centering
\includegraphics[width=0.6\textwidth]{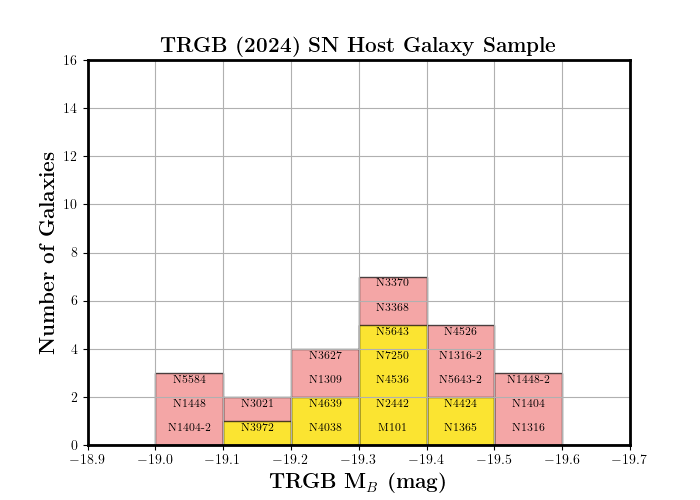}
 \caption{Histogram of absolute magnitude (M$_B$) values for 24 \sne in host galaxies for which CCHP TRGB distances have been measured. The squares in yellow indicate where 10 measurements have been made with \jwst. The full sample of 24 \sn increases the range of M$_B$, but the two overall distributions remain consistent. This larger sample of TRGB measurements can be compared to the M$_B$ distribution based on R22 Cepheids in Figure \ref{fig:MBhistSH0ES}, and again the two distributions are broadly consistent.   }
 \label{fig:MBhistCSP}
\end{figure*}

For comparison, we show in Figure \ref{fig:MBhistSH0ES} a histogram of the absolute magnitude (M$_B$) values for the host-\sne calibrating galaxies  in the \shoes sample. The Cepheid distances and apparent magnitudes (m$_B$) values are again from R22. The first point to note from Figures \ref{fig:MBhistCSP} and \ref{fig:MBhistSH0ES}  is their overall similarity. The range of M$_B$ values is comparable, and both exhibit central peaks.  The two histograms have been plotted to the same scale.  Shown in yellow are the 12 \sne that have the highest weight in the calibration. {\it These 12 \sne alone contribute half the weight of the \shoes sample.}

 \begin{figure*} 
 \centering
\includegraphics[width=0.6\textwidth]{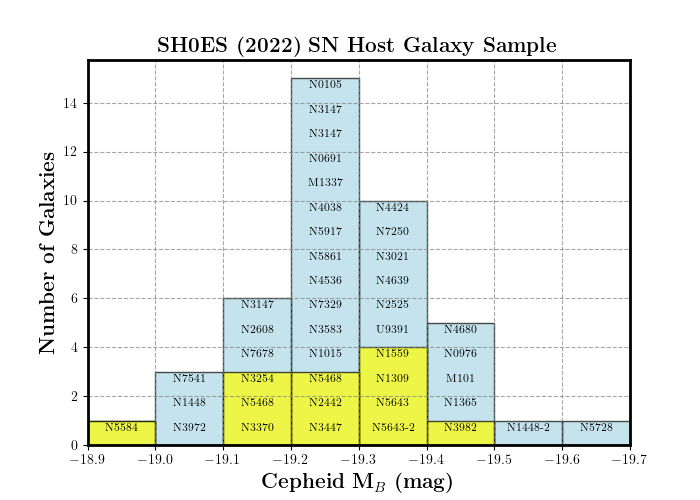}
 \caption{Histogram of absolute magnitude (M$_B$) values for 42 \sne in host galaxies for which R22 Cepheid distances have been measured. The squares in yellow indicate the 12 galaxies hosting the \sne that contribute half the weight of the \shoes sample. The distribution of \shoes M$_B$ magnitudes is broadly similar to that for the \cchp TRGB distribution. Of note, \ngc 5584 hosts SN 2007af, the faintest \sn in the \shoes sample, and the \sn with the highest weight. Removing this supernova from the sample decreases the dispersion in M$_B$ from 0.127 to 0.115 mag, and the value of \ho by 0.6 \hounits (half of the R22 \ho uncertainty).    }
 \label{fig:MBhistSH0ES}
\end{figure*}

In Figure \ref{fig:MBhistCSPcompare} we plot the M$_B$ histograms for the \csp distant sample of \sne and overplot (for a value of \ho = 70 \hounits) the nearby TRGB calibrators, this time on the \csp magnitude scale defined in \S \ref{sec:jwstcsp} above, and scaled by a factor of ten.

In terms of overall shape, this figure illustrates that the nearby TRGB \sn sample is a very reasonable match to the more distant, and larger \sn flow sample.  A Kolmogorov-Smirnov (KS) test comparing the two distributions yields a p-value of 0.12, indicating that the two distributions are consistent in shape. The same conclusion holds if we normalize the two histograms to the same area, and shift the nearby calibration sample by $\pm$ 0.03 mag (or 1\% in the \shoes value of \ho). 
Thus, we conclude that the TRGB sample is not biased (or skewed) with respect to the more distant sample, and it can provide a reliable calibration for the determination of \ho. 

\begin{figure*} 
 \centering
\includegraphics[width=0.6\textwidth]{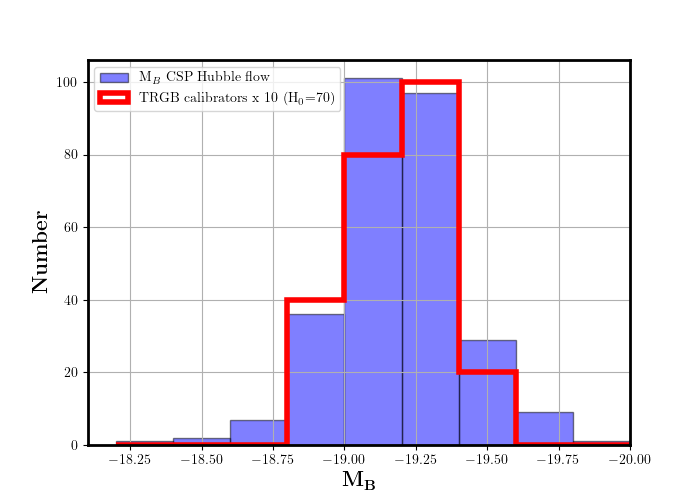}
 \caption{Histogram (in blue) of absolute magnitude (M$_B$) values for 291 \sne in the \csp sample. The red outline denotes the M$_B$ histogram for the TRGB calibration sample, scaled by a factor of 10. A KS test gives p=0.12, indicating that the two distributions are consistent. The comparison is intended to show only that the shapes are consistent.  }
 \label{fig:MBhistCSPcompare}
\end{figure*}

\section{Uncertainties in the M$_B$ Calibration for \sne}
\label{sec:sncalsys}

Achieving a 1\% accuracy in \ho measurements presents considerable challenges. Several concerns within the current M$_B$ calibration must be addressed before the true uncertainty in the local \ho value can be confirmed. A significant uncertainty stems from the limited number of local supernovae accessible to \hst and \jwst for Cepheid, TRGB, or JAGB distance measurements. Here we examine several issues impacting the accuracy of the current \ho value derived from the calibration of \sne, and assess the suitability of the existing calibrator sample for a 1\% accuracy estimate. Our conclusion is that claims of such precision are currently premature.

\noindent
1) At present, there are 42 \sne located in 37 galaxies where Cepheids have been discovered with \hst by the \shoes team.  For the closest \sne ($<$23 Mpc) sample with the smallest errors ($<$ 0.15 mag) the weighted mean M$_B$ magnitude is 0.08 mag (3.8\%) brighter than that for the rest of the sample (-19.32 $\pm$ 0.03 versus -19.24 $\pm$ 0.03 mag). While the samples are small and the difference amounts to approximately 2$\sigma$, this difference is large enough to warrant future study to determine if it is due to random statistical variation or an underlying systematic effect. Accurately constraining the systematic uncertainties is crucial. Even a subtle overall 0.03 mag difference corresponds to 1 \hounits, the entire size of the \shoes published uncertainty. Therefore, increasing the sample size and improving the statistics remain a key area for future follow-up.

As an example of how potential systematics can impact the measurement of Cepheid distances, in a detailed recent analysis of the period distributions of Cepheids in the \shoes sample, \citet{hogras_mortsell_2024} found that anchor galaxies exhibit a shorter period range compared to \sn host galaxies. Furthermore, galaxies with shorter period ranges, more closely resembling the anchors, yielded significantly lower \ho values. They note that this effect is consistent with a correlation between \sn absolute magnitude and distance modulus shown  in Appendix \ref{App:AppendixB} since shorter-period (fainter) Cepheids aresystematically undiscovered because of photometric incompletenessin galaxies at increasingly larger distances.

\noindent
2) The lower signal-to-noise, as well as the poorer resolution of the more distant sample may be contributing a systematic uncertainty.  For a desired total \ho uncertainty of 1\%, it is well to keep in mind that at distances greater than 40~Mpc, the Cepheids are crowded by stars within an area 33 times larger than at 7  Mpc (the distance to the calibrating galaxy, \ngc 4258) while, simultaneously the signal-to-noise is lower for the more distant sample. (At distances greater than 20~Mpc, Cepheids are crowded by stars contained within an area at least 8 times larger than the equivalent resolution element (fitting radius) at 7~Mpc.)  Future data will be required to assess  whether the more distant sample of Cepheid galaxies are free from systematic crowding bias or whether lower signal-to-noise results in increased uncertainties in the photometry of more distant galaxies. 

R24   have argued that crowding effects have been ruled out at an 8.2-$\sigma$ level of confidence. However, of the five galaxies for which they compare \hst and \jwst distance moduli, one  is extremely nearby (\ngc 5643 at 13 Mpc), two  (\ngc 1448 and \ngc 5584) showed the smallest differences between their 2016 and 2022 distance moduli (0.01 - 0.02 magnitudes) and thus are the least likely to have significant crowding issues,  and only one out of the five galaxies,  \ngc 5468, is at a distance of $\sim$40 Mpc. R24 assert that at 40 Mpc one would expect a $\sim$0.30 mag effect to explain the \ho tension due to crowding. (This of course would hold only if {\it all} of the tension were to be explained by crowding, a highly unlikely explanation.) However, it should be noted that a 0.03 mag systematic error (i.e., comparable to what was found in the nearby sample) would result in a systematic error in \ho of 1 \hounits, the entire size of the R22 \ho uncertainty. As noted, most of the (most challenging distant) galaxies beyond $\sim$ 40 Mpc have not yet been imaged at the improved resolution of  \jwst.

\noindent
3) Some of the challenges in measuring accurate distances to the nearest galaxies in this sample can be seen in the Figure \ref{fig:muR16R22}. Shown is a comparison of  the  distance moduli for \sne calibrator galaxies in common to the R16 and R22 studies (based on distances published in their Tables 5 and 6, respectively). The \hst data for these original 19 galaxies (20 \sne) were first analyzed in R16, and the same raw data were  then re-analyzed and re-calibrated in R22. Differences up to 0.3 magnitudes in distance modulus were measured, with a net increase in the mean distance modulus (R22-R16) of 0.035 $\pm$ 0.025 mag (and a brightening of  M$_B$ by the same amount). Considered in isolation, this  systematic difference would result in a decrease in \ho by 1.6\%. Indeed, if one retroactively corrects the R16 distances to the R22 scale correcting for the 0.035 mag offset,  the \shoes \ho value becomes 72.05 $\pm$ 1.74 \hounits, resulting in a lower (2.6$\sigma$) and much less interesting Hubble tension (see also \S\ref{sec:tension}).

\begin{figure*} 
 \centering
\includegraphics[width=1.0\textwidth]{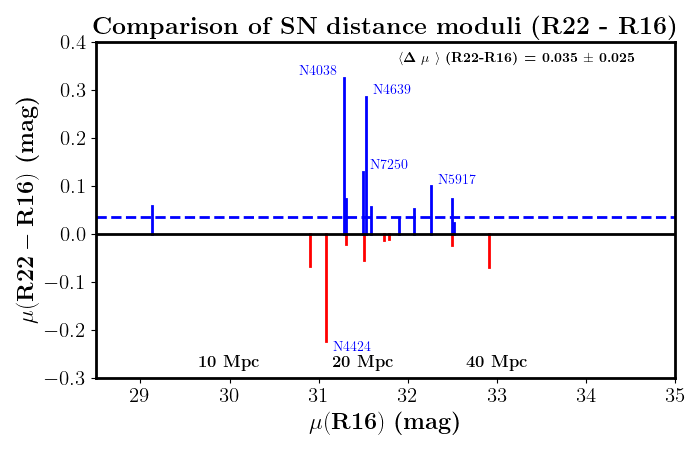}
 \caption{  Comparison of distance moduli measured by \shoes in R16 and reanalyzed in R22.  The galaxies with the largest shifts are labeled. There are differences up to 0.3 magnitudes in the calibrating distance modulus values, with  an increase in the mean value of $\mu$ of 0.035 $\pm$ 0.033 mag, shown by the horizontal dashed blue line.  Based on this difference alone, the R22 calibration would result in a decrease in \ho of 1.6\% compared to R16. }
 \label{fig:muR16R22}
\end{figure*} 

\noindent
It must be noted that 60\% of the R22 sample galaxies are located at distances $>$20 Mpc and 25\% are more than 40 Mpc distant. The more distant galaxies, on average, have lower signal to noise (S/N) and decreased resolution. Fortunately, this is an issue that can be resolved with further higher resolution \jwst data for the more distant sample, and it is clearly warranted before crowding can be definitively ruled out as a potential systematic.

\noindent
4) The photometric calibration of \sne is still undergoing changes with systematic shifts the size of the} \shoes total error bar.  In Figure \ref{fig:MBvsDelta_mb}, we show the change in adopted apparent \sn magnitudes ($\Delta$m$_b$) between the \citet{riess_2016} and \citet{riess_2022} studies  plotted as a function of M$_B$. The point sizes scale with the weight, w,  for each object in the M$_B$ calibration, as defined by 

\begin{equation}
\label{eq:weights}
w_i = {1 \over \sqrt{\sigma_{\mu,i}^2 + \sigma_{m_B,i}^2
} }
\end{equation}

\noindent
The shift in apparent magnitude is systematic: two thirds (13 of the 19) of the \sne have fainter magnitudes in the R22 study.  The mean shift in apparent magnitude amounts to 0.03 mag, partially compensating for the decrease in \ho that would result from the increase in the \shoes mean distance moduli between R16 and R22 (see \S \ref{sec:compareTRGBwithR22}).  

\begin{figure*} 
 \centering
\includegraphics[width=0.6\textwidth]{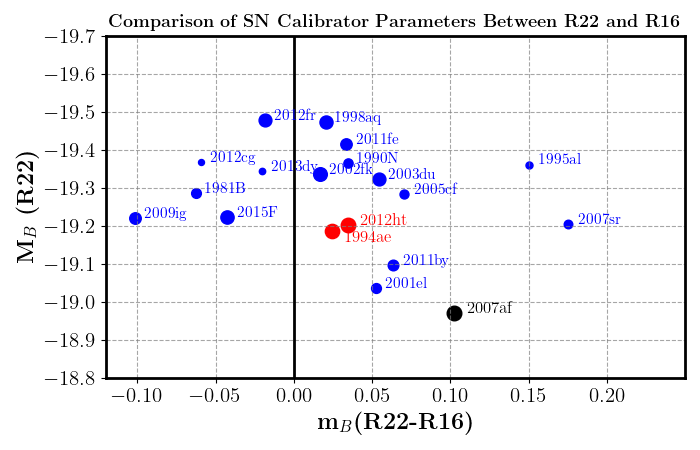}
 \caption{M${_B}$ values from R22 versus the difference in apparent \sn magnitude adopted between  2016 and 2022, labeled as m$_B$ (R22-R16). The size of the points is proportional to the weight for each \sn in the calibration, as defined in the text. Most of the \sne in the sample (13/19) have adopted  m$_B$ values in 2022 that are larger (fainter) than those adopted in 2016. The three \sne with the highest weight in this common sample of 19 objects are SN 2007af (black circle, the highest weight in the sample) and SN 2012ht  and 1994ae (both in red). All three of these \sne have absolute magnitudes values falling toward the faint end of the M$_B$ distribution. The mean shift in m$_B$ from 2016 to 2022 is +0.03 mag or 1.4\%, the current size of the total R22 \ho uncertainty. }
 \label{fig:MBvsDelta_mb}
\end{figure*}

\noindent
5) The faintest supernova in the \citet{riess_2022} sample is SN2007af in the galaxy \ngc 5584. From Figure \ref{fig:MBhistSH0ES}, this supernova can be seen at the faint end of the distribution. For the nearest half of the sample (comparable in distance to our \jwst sample), SN 2007af is the most discrepant point in the weighted average. In this nearest half of the sample, the weighted mean $M_B$ = -19.284 $\pm$0.039~mag (error of the mean). Removing the single supernova, SN 2007af increases the weighted mean magnitude to M$_B$ = -19.316 $\pm$ 0.032 mag. (This value can be compared to the  value of M$_B$ = -19.317 $\pm$ 0.033  mag that is obtained for the nearby \jwst sample using the \shoes data.)

In comparing 17 \sne in common between the \shoes and \cchp studies, H25 notes how the pull of outliers is significantly reduced in the TRGB calibration. For this overlapping sample, recall that in the case of the \shoes analysis, if SN2007af is dropped from the sample, M$_B$ shifts by 0.029 mag (1.3\%), corresponding to a shift in \ho of -1.02 \hounits.  Note that this shift, due to the exclusion of a single \sn calibrator, is as large as the entire R22 uncertainty in \ho.   However, this is not the case for the \cchp TRGB calibration, where M$_B$ shifts by a negligible amount of only 0.004 mag, or -0.15 \hounits.

In the case of the total sample of 42 \sne, dropping SN2007af  results in a decrease in \ho of 0.5 \hounits (half of the R22 \ho uncertainty) and it lowers the dispersion in the weighted mean from 0.129 to 0.115 mag. We emphasize again that a major advantage of the TRGB distance scale is the ability to  measure stars in the outer regions of galaxies where  dust and crowding effects are minimal. The errors in the TRGB distance moduli range from 0.04 to 0.10 mag (a factor of 2.5), whereas for Cepheids, the uncertainties differ by an order of magnitude, ranging from 0.037 to 0.253 mag (a factor of 6.8); i.e., spanning nearly a factor of 50 in the variance (and therefore strongly weighting different supernova disproportionately,  with some being 50 times more influential than others) in the calibration.

\noindent
6) H25 discusses at length the \cchp TRGB and \shoes Cepheid calibrations of the \cspi + \cspii, as well as SuperCal \citep{scolnic_2015} and Pantheon+ \citep{scolnic_2023} \sn data. Restricting the comparison of \csp and Pantheon+ to the same overlapping subset of \sne (22 in the case of \cspi and 14 for SuperCal, H25 finds that in the case of the \csp data, the larger \hst + \jwst sample retains a stable central \ho value when compared to the \jwst sample only, differing by only -0.11 \hounits. However, this does not occur in the case of the SuperCal and Pantheon+ data.  In the latter case, the difference in \ho between the SuperCal and Pantheon+ data is +1.20 \hounits. Note that this difference cannot be explained by differences in the TRGB and Cepheid distances (which now agree to $\sim1\%$, see \S \ref{sec:compareTRGBwithR22}). This difference, once again greater than the total R22 quoted \ho uncertainty, is now largely due to changes in the apparent magnitudes, m$_B$, introduced in the transition from SuperCal to Pantheon+ (e.g., see Figure \ref{fig:MBvsDelta_mb}). We note also that six of the eight calibrations discussed by H25 yield values of \ho $<$ 70 \hounits. A single calibration gives \ho $>$ 71 \hounits. The Pantheon+ calibration is a significant outlier in the \hst + \jwst TRGB calibration, shifting the Hubble constant by +1.7 \hounits.

\noindent
7) Half of the weight in the \shoes sample of M$_B$ values comes from only a dozen \sne. These 12 \sne represent only 29\% of the total sample of 42. An already small sample of 42 \sne is not providing $1/\sqrt{N}$ statistical (increased sample size) gains. An effective sample size can be computed as follows: 
\begin{equation}
N_{\text{eff}} = \frac{(\sum_{i=1}^{42} w_i)^2}{\sum_{i=1}^{42} w_i^2} = 31
\label{eq:neff}
\end{equation}

\noindent
That is, the effective size of the R22 Cepheid sample is equivalent to only 31 \sne, or 74\% of the total sample. (The effective size of the \cchp TRGB sample is 21 compared to a total of 24, or 88\%; i.e., the weights are more evenly distributed.) These limitations, however, underscore the need for a larger sample of nearby \sne. 

In summary, recent changes to  the Cepheid distance  scale are illustrative of the types of potential (at-the-time-unknown) systematic errors, having sizes comparable to the total current estimated uncertainties.    Furthermore, recent revisions to the originally published apparent magnitudes of these supernovae, comparable in scale to the entire quoted \shoes uncertainty, suggest that the \sn calibration may not have yet converged.   Despite the \shoes program's use of 42 SNe Ia calibrated with Cepheid distances, the current Cepheid distance and \sn calibration uncertainties  result in less than one-third of the supernova sample (12 objects) contributing half the overall calibration weight.  The implications for a claim for physics beyond the standard model clearly demand a larger sample of nearby \sne on which to base the calibration. Even a subtle difference, such as a 0.03 magnitude variation, can significantly impact the value of \ho at a level of the  current quoted error budget.  Below we review the tension in \ho, and conclude that the current state of the nearby calibration does not support a tension level of 5$\sigma$.

\section{The Tension}
\label{sec:tension}

Given that significant systematic offsets are still being identified empirically that are comparable to the current total estimated  1$\sigma$ uncertainty indicates that representative or realistic uncertainties have not been previously accounted for in estimating the true tension.  The claim of 5$\sigma$ tension does not allow for the observed systematic evolution of the measurements, which violates the assumption of statistical independence and likely inflates the significance.

A tension of 5$\sigma$ follows from absorbing the uncertainties in \ho into a single quantity as follows: 

\[
T = \frac{\left| H_0^{(R22)} - H_0^{(Planck)} \right|}{\sqrt{\sigma_{R22}^2 + \sigma_{Planck}^2}}
\]

\noindent
where $H_0^{(R22)} = 73.04 \pm$ 1.04 and $H_0^{(Planck)} = 67.4 \pm $ 0.5 \hounits.

As an illustrative example, given here to evaluate the impact of these shifts on the calculated Hubble tension, we consider several recent changes to the \shoes Cepheid distances. We compare the distances in the (baseline) R22 Cepheid sample with those obtained three years previously in the \citet{riess_2019} [hereaftr, R19] study.  R19 did not tabulate their distance moduli, but the average difference between the R22 and R19 Cepheid distance moduli amounts to $+0.058$ mag (which is almost a 3\% change expressed in  distance).
This difference reflects:
\begin{enumerate}
    \item The difference between R19 and R16, where R16 values were updated as follows:
    \begin{itemize}
        \item An update to the LMC distance from \citet{pietrzynski_2013} to \citet{pietrzynski_2019} and the recalibration of \hstwfciii-IR LMC Cepheid photometry amounting to a total of R19-R16 = -0.029 mag. This update resulted in a value of \ho = 74.22 $\pm$ 1.82 \hounits based on the LMC alone (differing from that of \ho = 72.04 $\pm$ 2.67 in R16). Combining the LMC, Milky Way and \ngc 4258 distances resulted in a value of \ho = 74.03 $\pm$ 1.42 (corresponding to a difference of only +0.006 mag from the LMC alone). 
    \end{itemize}
    \item The average difference between the \citet{riess_2022} and \cite{riess_2016} published distance moduli (where R22-R16 = +0.035 mag) 
\end{enumerate}

The average difference between R22 and R19 thus amounts to: 
\par\noindent[(R22-R16) - (R19-R16)] = [(+0.035) - (-0.029 + 0.006)] = +0.058 mag.

We first subtract the systematic shift in \ho (2.00 \hounits) due to the revision to the R19 Cepheid distances (i.e., the impact on \ho due to the difference of +0.058 mag in the average distance moduli). We take the recent concordance in the \shoes Cepheid and \cchp TRGB distances as an indication that this correction is reasonably robust.  Given the fact that the \sn calibration has not yet converged among the various projects \citep[see also][] {DESI_2024}, we then take the difference in the apparent \sn magnitudes (the shift of 0.03 mag) as an indication of outstanding systematic uncertainties in the \sn calibration (and also reflective of the difference in the Pantheon+ and CSP plus SuperCal (see H25) \sn data sets). For the purpose of this exercise, we retain the small 1\% uncertainty quoted by R22 and add in quadrature the uncertainty in the \sn calibration. We then recalculate the tension as follows:  

\[
T = \frac{\left| H_0,corr^{(R22)} - H_0^{(Planck)} \right|}{\sqrt{\sigma_{R22,corr}^2 + \sigma_{Planck}^2}}
\]

\noindent
where $H_0, corr^{(R22)}$ = 74.03 - 2.00 = 72.03 \hounits;  $\sigma_{SN}$ = 1.00 in \hounits, and $\sigma_{R22,corr}$ = $\sqrt{\sigma_{R22}^2+\sigma_{SN}^2}$ = 1.44.

When including this systematic uncertainty in the supernova calibration, the tension is reduced to 3.0$\sigma$ , and is  reflecting a more  conservative assessment of the measurement's reliability. Note also that this simple calculation assumes an uncertainty of $\sigma_{R22}$ = 1.04, which may still be an underestimate. (Adopting the error on the R19 \ho value of $\sigma_{R19}$ = 1.42 results in a tension of only 2.6$\sigma$.) In their discussion of  additional potential, unaccounted-for uncertainties,  \citet{hogras_mortsell_2024}  independently conclude that differences in the period distributions of the anchor relative to host galaxies results in only a 2.4$\sigma$ tension (when using the Milky Way zero point calibration). 

We conclude that considerably more caution is warranted before concluding that the Hubble tension is at the level of 5$\sigma$. A 5$\sigma$ tension implies that there is only a 0.00003\% probability (a vanishingly small number where astrophysical measurements are concerned) that the observed difference is due to random fluctuations. We note also that a 5$\sigma$ tension is only meaningful for a normal distribution. Claiming that the $\Lambda$CDM model is incomplete carries with it the implicit assumption that it is the local value of \ho that is the correct value, holding the standard model as the one in need of change.  However, the tension itself carries no information on which value is in need of updating. We have already seen how the \shoes distance scale has shifted by nearly 1.6\%, and the \sn magnitudes have shifted by 1.4\%,  both systematic differences comparable in size to the R22 total 1.4\% quoted uncertainty. While it is not possible to assess the size of unknown systematics in advance, it is possible to examine recent prior estimates with an eye to previously hidden systematic uncertainties.  While differences of order the size of the total quoted uncertainty are still at play, 
a claim of 5$\sigma$ tension may be premature.

\medskip\medskip\medskip\medskip\medskip\medskip\medskip\medskip\medskip\medskip
 
\section{Error Budget}
\label{sec:errors}

\subsection{Distance Errors}

The statistical and systematic uncertainties for each of the three distance indicators are discussed in detail in the companion papers of H25 and L25. 
Here we note that in the determination of \ho, the overall uncertainties for the individual galaxy distances become statistical uncertainties. For example, while the aperture corrections and  reddening corrections for an individual galaxy contribute to its systematic uncertainty, for an ensemble of galaxies, the aperture corrections  and/or reddening corrections are uncorrelated from galaxy to galaxy, and become a source of random/statistical uncertainty. The overall statistical uncertainty is determined by the number of host galaxy calibrators, and the total number of \sne contained therein.

In Table \ref{tab:hoerrors} we summarize the sources of the statistical and systematic uncertainties in our \ho measurement. We list the statistical and systematic errors separately. 
The individual statistical uncertainties were discussed in \S \ref{sec:mcmc}.) 
The systematic error common to both the TRGB and JAGB methods is that for the zero point (anchored to \ngc 4258). As discussed in \S\ref{sec:comparison}, there is an independent systematic uncertainty for each method that results from the measurement uncertainty in the TRGB or JAGB magnitude for \ngc 4258. These two errors are added in quadrature and given in Table \ref{tab:hoerrors}. Finally, we introduce an additional 1\% error term, denoted as $\sigma_{SN}$, to account for residual uncertainties in the supernova calibration, including photometric calibration of M$_B$ and peculiar motion effects.

\subsection{Supernova Errors}

For \sn observations, a common limitation of all \ho measurements (\shoes/Pantheon+ and \cchp/\csp alike) is that they rely on data collected using various telescopes, instruments, and filter combinations, making the sample of nearby host \sn galaxies heterogeneous. However, for distant \sne, the \csp has an advantage, as all observations were conducted at Las Campanas using consistent filter combinations and detectors. In some cases \csp data were obtained for both nearby and distant \sn samples, in which case the error in zero point cancels.  However, since the \csp did not begin until 2004, any \sne discovered earlier lack \csp observations. A similar limitation applies to \sne in northern hemisphere galaxies that were inaccessible to southern telescopes at Las Campanas (e.g., 2011fe in M101). The typical error in optical zero points is around 0.01 mag; however, to be conservative, we adopt a systematic uncertainty of 0.015 mag to account for the inhomogeneous nature of the current calibration.

All of the nuisance parameters in the \sn magnitudes, as characterized in Equation \ref{eq:m_Tripp}, are parameters in the MCMC analysis, and are  included in the final uncertainty in \ho from that analysis. The \csp K-corrections are all computed with the same time series spectral energy distribution template, which could result in a systematic error. However, as can be seen from \citet{lu_2023}, Figure 16,    the  systematic uncertainties in the near-infrared K-corrections are,  in any case, very small ($<<$1\%); the optical data have more spectra contributing to the template, and their uncertainties are also very small.

\subsection{Adopted Uncertainty on H$_0$}

We adopt the uncertainties as described above in \S \ref{sec:errors} and shown in Table \ref{tab:hoerrors}.
Based on our \hst + \jwst TRGB sample of 24 \sne, as described in \S \ref{sec:mcmc} and illustrated in Figure \ref{fig:corner},
our adopted value of the Hubble constant and its uncertainty, as derived from our new \jwst distances applied to the \csp sample of \sne is \ho = 70.39  $\pm$ 1.22 (stat) $\pm$ 1.33 (sys) $\pm$ 0.70 [SN] \hounits.

\begin{deluxetable*}{lcccc}
\tablecaption{Summary of \ho Uncertainties \label{tab:hoerrors}} 
\tablehead{\colhead{Source of Error} & \colhead{Random Error} & \colhead{Systematic Error} & $\sigma_{SN}$ &\colhead{Reference}} 
\startdata
TRGB (\jwst)          &       2.6\%          &       1.9\%\tablenotemark{a}  &  1\%          &   H25, \S\ref{sec:prevTRGBJAGB},\S\ref{sec:mcmc}, this section           \\
JAGB (\jwst)           &       3.0\%         &       2.4\%\tablenotemark{b} & 1\%             &   L25, \S\ref{sec:prevTRGBJAGB}, \S\ref{sec:mcmc}, this section         \\
Combined distances           &       2.4\% \tablenotemark{c}   &       1.8\%   &1\%          &  \S\ref{sec:prevTRGBJAGB},\S\ref{sec:mcmc}  \\
\hline  \hline
TRGB (\hst + \jwst)           &       1.7\%   &       1.9\%             &  1\% & \S\ref{sec:mcmc}, this section  \\
\enddata
\tablenotetext{}{(a) Uncertainty of 1.5\% in \ngc 4258 distance; \citet{reid_2019} and \ngc 4258 TRGB measurement uncertainty}
\tablenotetext{}{(b) Uncertainty of 1.5\% in \ngc 4258 distance; \citet{reid_2019} and \ngc 4258 TRGB measurement uncertainty}
\tablenotetext{}{(c) TRGB and JAGB combined distances and errors }
\end{deluxetable*}

\subsection{Overall Systematic Uncertainties}
\label{sec:overall_systematics}

We describe here the systematic effects that are common to the local distance (TRGB, JAGB, Cepheid and \sn) scales. Although generally small,  in an era of accurate cosmology, percent-level uncertainties will become more significant.

\begin{enumerate}

\item The calibrations of the JAGB stars and TRGB   for the \jwst data presented in this paper are both  anchored to the geometric distance to one galaxy, \ngc~4258, and  these two methods therefore share any systematic errors (known or unknown) in that determination. The currently cited total uncertainty in the \ngc 4258 distance is 1.5\% \citep{reid_2019}. However, the TRGB distances measured with \hst \citep{freedman_2021}, and tied to four geometric anchors, agree to within 1\% of the \jwst TRGB calibration based on \ngc 4258 alone.  
In forthcoming papers, we will continue to improve upon and apply additional  zero-point anchors to the TRGB, JAGB and Cepheid distance scales.

\item As discussed in \S \ref{sec:sncalsys}, a current key source of uncertainty in the \sn distance scale is the relatively small number of well-observed \sne with calibrators that have distances measured using \jwst. One concern is a trend (see Figure \ref{fig:MBdistR22}) in which \sn absolute magnitudes appear to become fainter with increasing \hst-measured Cepheid distances (and increasing errors). This trend could arise from statistical effects or systematic issues in the measured distances, particularly in galaxies where crowding/blending effects, lower signal-to-noise, or differences in the period ranges covered may introduce biases.

\item We have adopted the same reddening law, which is assumed to be universal, with the same ratio of total-to-selective absorption, $R_V$. For the JAGB and TRGB, where foreground reddening corrections are required, there is an uncertainty due to the zero-point calibration of the extinction law. \citet{schlegel_1998} estimates this uncertainty to be $E(B-V)$ = 0.02 mag, which corresponds to an extinction uncertainty of A$_J$ = 0.016 mag (0.7\%).

\item In this analysis, we use \sne to extend the distance range for the \ho determination. Any remaining systematic error intrinsic to \sne will be shared by all of the local distance methods. These systematics might include inconsistent calibrations 
of photometric zero points across different surveys and instruments that can lead to systematic offsets ({\it e.g.,} in the case of Pantheon+); in addition the calibrator light curves are generally not part of the well-calibrated \csp survey, which makes up the Hubble flow sample. Further uncertainties may arise from dust extinction corrections, including the possibility of multiple dust types (e.g., in the Milky Way, the host galaxies or dust surrounding the \sn) and potential differences in the extinction law). Other sources of uncertainty include intrinsic scatter in \sne peak luminosities and effects such as the host galaxy mass step, which may not be fully constrained. These factors could introduce residual, uncorrected biases in anchoring the luminosities of calibrator \sne to those of more distant \sne. 

\item Finally, the results of an MCMC analysis may not be as robust as the confidence levels implied by the likelihood analysis. These include the fact that the probability might be accurately treated as a Gaussian near the peak, but it may not decrease as fast as a Gaussian in the tails; there may be unknown systematic errors  revealed as the statistical uncertainty  decreases and a systematic floor is reached; and the likelihood function may have uncertainties and/or assumptions that have not been included in the quoted uncertainty \citep[e.g.,][]{liddle_2009, hogg_foreman-mackey_2018}.
\end{enumerate}

\section{Comparisons With Previously Published Data}
\label{sec:previous}

\subsection{Comparison of  TRGB and $I$-band Distances}
\label{sec:prevTRGBJAGB}

\begin{figure}
\centering
\includegraphics[width=15.0cm, angle=-0]
{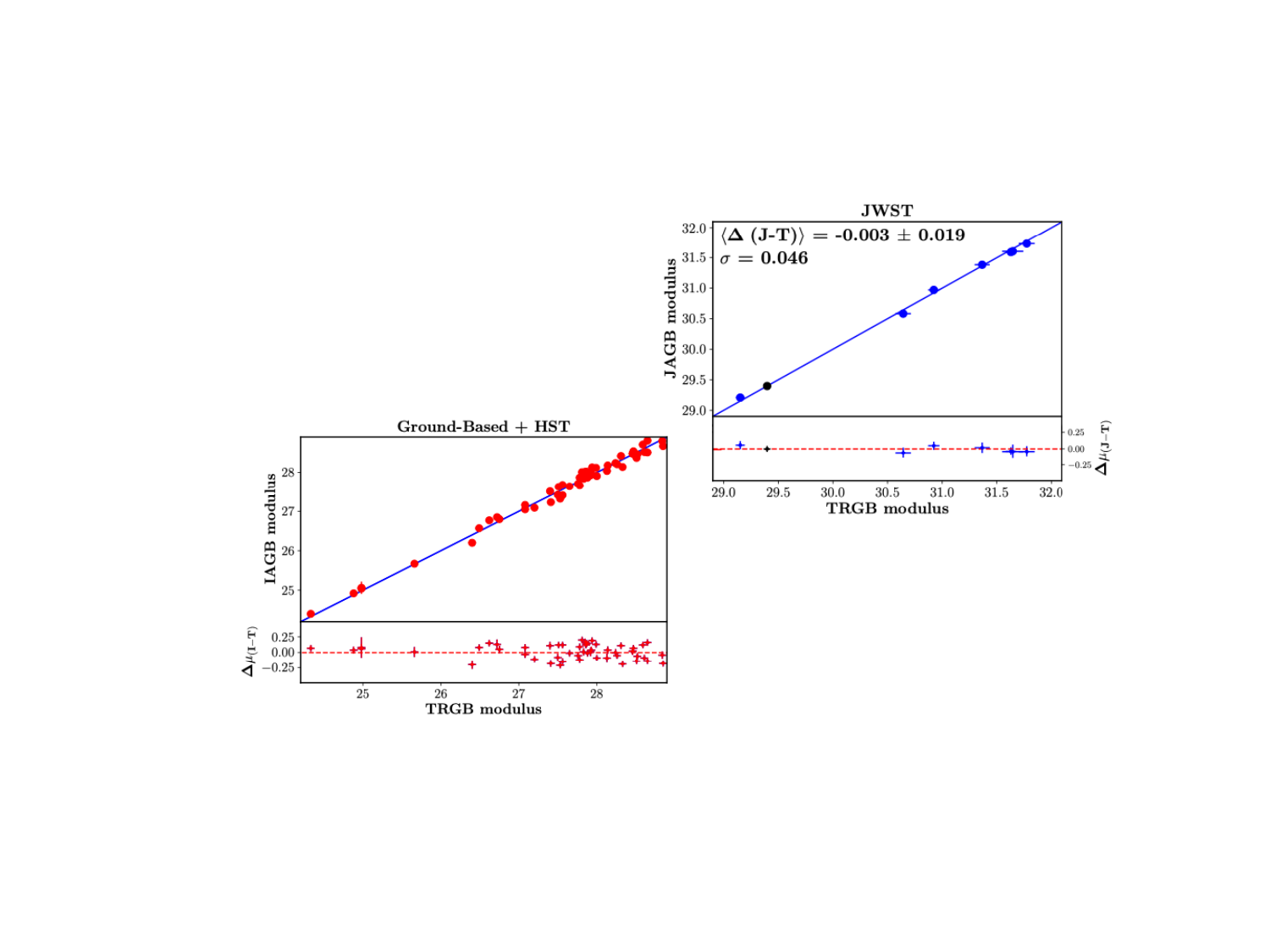}
\caption{Left panel:  Comparison of I-band AGB/carbon and TRGB distances for  ground-based and \hst observations from \citet{freedman_2025} (red dots).  Right panel: Our new \jwst distances (blue dots). Residuals are shown beneath each plot. There is excellent agreement in the distances spanning 50 kpc to 23 Mpc. 
\label{fig:compareGd_JT_CT}}
\end{figure}

In Figure \ref{fig:compareGd_JT_CT}, we compare measured TRGB distances with those from AGB/carbon stars measured in the $I$-band, including nearby galaxies measured with ground-based telescopes and \hst, and tabulated in \citet{freedman_2025}. 
There is strong consistency, with remarkably low scatter among these two  distance scales, over a range of a factor of 50 in distance.  These comparisons provide an additional external constraint on both the zero-point calibrations of the TRGB and JAGB methods, as well as their respective internal precisions. 

\subsection{Comparison of H$_0$ Values: $I$-band TRGB Distances Measured Using \hst}
\label{sec:trgbcompare}
 
In \citet{freedman_2019, freedman_2020, freedman_2021} we presented results from a \hst program to measure $I$-band TRGB distances based on observations of 20 \sne located in a sample of nearby galaxies. In \citet{freedman_2021} four geometric anchor distances were applied to the $I-$band TRGB distance scale: (1) \gaia parallaxes for Milky Way globular clusters \citep{cerny_2021a, maiz_apellaniz_2021, vasiliev_baumgardt_2021}; (2) the detached eclipsing binary distance (DEB) to the LMC \citep{hoyt_2023a, pietrzynski_2019}; (3) the DEB distance to the SMC \citep{hoyt_2023a, graczyk_2020} and (4) the maser distance to \ngc 4258 \citep{reid_2019}. Using these four geometric anchors, the calibration yielded \ho = 69.8 $\pm$ 0.6 (stat) $\pm$ 1.6 [sys] \hounits, a result that is in excellent statistical agreement with the \jwst analysis presented in this study.

Of the 20 \sne in \citet{freedman_2021} and the 10 galaxies in the current study with \jwst TRGB distances, there are six galaxies in common to both studies: M101, \ngc 1365, \ngc 4038, \ngc 4424, \ngc 4536 and \ngc 5643.  The differences in distance moduli (in the sense \jwst $J$-band minus \hst $I$-band) are +0.073,  0.006, -0.036, -0.074, -0.041, and  +0.167, respectively, with a weighted average distance modulus offset amounting to only +0.008 $\pm$ 0.030~mag (error on the mean). 

We note that the data for these two studies are entirely independent: the $I$-band data for the target galaxies were obtained with \hst, whereas the $J$-band data were obtained with \jwst. Furthermore, and importantly, the $I$-band distances were measured with respect to the LMC, the SMC, the Milky Way and \ngc 4258, whereas the $J$-band distances were measured with respect to \ngc 4258 alone.  This independent comparison serves to provide a strong external check on the TRGB distance scale.

\subsection{The Run of H$_0$ over the Past 25 Years}
\label{sec:howithtime}

\begin{figure}[ht!]
\plotone{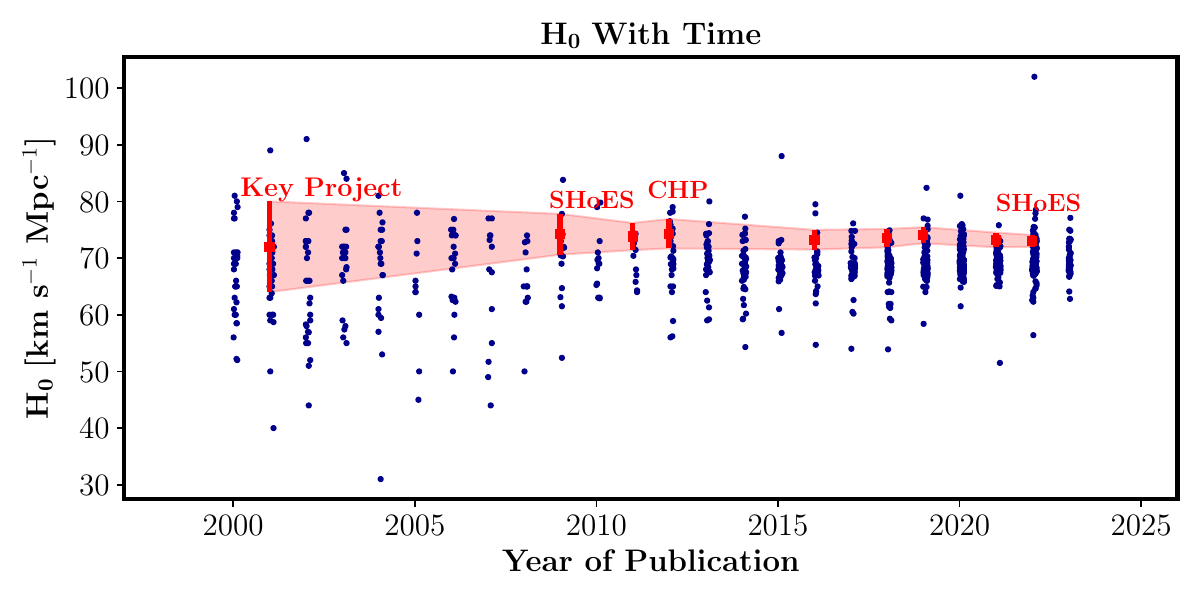}
\caption{Individual values of \ho (blue dots) as a function of publication date, using the compilation of \cite{steer_2020}, updated by Steer (2024, priv. comm.) shown. The larger red filled circles (with vertical error bars) are \ho values calibrated using Cepheid distances. The light red shaded region tracks the one-sigma quoted uncertainties of the Cepheid \ho measurements, which, as can be seen, have decreased significantly over the past 25 years. }
\label{fig:ho_only_withtime}
\end{figure}

While both the precision and the accuracy of the measurements of distances to galaxies  have improved considerably in recent years, there are still reasons to remain open as to whether our currently estimated uncertainties truly reflect the total uncertainties.  Shown in  Figure \ref{fig:ho_only_withtime} are all published \ho values from the database maintained by Ian Steer (\citet{steer_2020}, and updated in 2024, priv. comm.; small dark blue dots) and the values of \ho calibrated by Cepheids (larger red filled circles).

\medskip\medskip\medskip

There are several points worth noting in Figure \ref{fig:ho_only_withtime}. First, the values of \ho obtained from \sne, as calibrated by Cepheids, have remained remarkably constant for over a quarter of a century. Second, the Cepheid-based values of \ho are systematically higher than the (mean, median or mode) of the distribution of other determinations (Steer et al. 2020, 2024). Third, there is no bimodal distribution of published values centered, respectively, around the oft-quoted values of 67 and 73 km/s/Mpc. Future studies will be required to unambiguously confirm the higher \ho values that are based on Cepheid measurements alone, or determine whether systematic errors ultimately prove to be the explanation.

\medskip\medskip\medskip\medskip\medskip\medskip\medskip\medskip\medskip

\section{Future Prospects}
\label{sec:future}

With JWST, we have the means of improving both the accuracy and the precision of the locally determined extragalactic distance scale, solidly providing a multiply calibrated and independently verified determination of the current expansion rate of the universe, \ho. 

\begin{enumerate}
\item{\it Improving the Zero-Point (Anchor) Calibration:} At present, there are four galaxies for which geometric distances can be measured for calibrating astrophysical/stellar distance indicators suitable for determining the extragalactic distance scale. These include the LMC, \ngc 4258, and the SMC, with quoted uncertainties of  
1\% \citep{pietrzynski_2019},  1.5\% \citep{reid_2019}, and 2\% in distance \citep{graczyk_2020}, respectively. The Milky Way parallax offset currently has a significant impact on derived parallaxes for Cepheids and the period-luminosity-metallicity relation \citep[e.g., see][for a discussion of the current uncertainties]{groenewegen_2021, owens_2022}. Future data releases for the \gaia satellite are forecast to provide a 1\% parallax calibration for the Cepheid PL relation. In addition, accurate parallax measurements for Milky Way globular clusters will provide a zero-point calibration for the TRGB, again at a level of 1\%.

\item{\it Increasing the Numbers of SN Ia Host Calibrators:} On average, only about one new \sn, per year,  is found within a distance accessible for follow-up discovery of Cepheids, TRGB or JAGB stars using \hst. Thus the total sample of \sn host galaxies is not expected to increase significantly over the next decade. At present there are 37 galaxies for which \hst Cepheid distances have been measured as part of the \shoes project. Ten supernova-host galaxies have been observed with \jwst as part of this paper (GO-1995), and an additional 11 have been observed with \jwst for programs GO-1685 and 2875. One galaxy is in common to both programs, \ngc 5643, and both programs additionally have observations of \ngc 4258.  \jwst GO-3055 includes observations in the outer regions of \ngc 1404 and \ngc 1380 with the filters $F090W$ and $F150W$, optimal for TRGB and JAGB measurements. 
In the case of \jwst, doubling the sample of galaxies observed will improve the precision of \ho by a factor of $1/\sqrt 2$. Beyond improving the statistical uncertainties alone, observations of more distant galaxies, where crowding effects are more severe, will be important for constraining potential systematic uncertainties increasingly encountered as a function of distance.

Currently planned 30-meter-class ground-based optical telescopes (e.g., the Giant Magellan Telescope) will have 10 times the resolution of \hst, and will allow Cepheids, TRGB and JAGB stars to be discovered and measured within a 1,000 times greater volume, thereby providing a significant increase in the numbers of \sn host galaxies.  The Cepheids can then be followed up in the near-infrared with \jwst or the Nancy Grace Roman Space Telescope (see below).

\item{\it Photometrically Consistent and Wider Areal Coverage of Nearby Galaxies:}  The Nancy Grace Roman Space Telescope\footnote{\url{https://roman.gsfc.nasa.gov/}}, due to be launched in 2027, will have 100 times the field of view (FOV) of \hst, with comparable resolution and sensitivity, with wavelength coverage  extending from 0.48 to 2.  
  3$\mu$m. The large FOV will enable the mapping of entire disks and halos of nearby \sne host galaxies, giving simultaneous measurements of the TRGB, Cepheids, and JAGB stars in single (massive) pointings. The simultaneous measurements will provide a consistent photometric calibration for all three methods. The European Space Agency's Euclid\footnote{\url{https://www.esa.int/Science_Exploration/Space_Science/Euclid_overview}} satellite has a wide field of view of 0.67 deg$^2$ and is already providing detailed, high-resolution CMDs of the nearest galaxies \citep{hunt_2024} out to $\lesssim$ 5 Mpc. Although the resolution is insufficient to measure \sn host galaxies directly, comparison of the distances measured using TRGB and JAGB stars will provide valuable tests of measurement systematics. 

\item{\it Future SN Surveys:} Over the last few decades,  an enormous amount of time and energy has gone into ground-based surveys for \sne (e.g., Pantheon+). This database is the merger of observations from 18 different surveys \citep{scolnic_2022}, taken with different telescopes, instruments, calibrations, etc. The Vera C. Rubin Observatory\footnote{\url{https://rubinobservatory.org/}}, which will begin taking data in 2025, is forecast to discover over 3 million supernovae in its first decade of operation. Follow-up spectroscopy to measure redshifts using other ground-based facilities will pave the way for a dataset that has a consistent photometric calibration, without the need to apply photometric offsets and/or corrections to heterogeneous samples, as is the case today. 

\item{\it Additional Tests for Systematics:} Importantly, future measurement of \ho employing techniques completely independent of the local distance scale,  accurate at the 1-2\% level, will be essential for ruling out  remaining systematic uncertainties in the local distance scale. Examples include gravitational wave sirens \citep[e.g.,][]{chen_fishbach_holz_2018} and gravitationally lensed supernovae \citep[e.g.,][]{grillo_2024}, which hold exciting promise if more objects can be discovered and their precision and accuracy improved.

\item{\it More Accurate Modeling to Standardize \sne: } A promising  path to increasing precision and accuracy in \ho, at once addressing Points 2 and 5, is improvements to models for standardizing \sn Hubble residuals \citep[e.g.][]{boone_2021,stein_2022}. For these models, the \sn Hubble residuals can be  standardized to 0.07-0.08~mag in distance using  time-series spectrophotometry (compared to typical dispersions in the 0.15~mag range for light curve standardization approaches). The smaller intrinsic dispersion of such models can both significantly increase the amount by which the statistical precision of \ho tightens per \sn calibrator and, perhaps more importantly,  also tighten the constraints on any as-yet unseen systematics in \sn standardization, which are more likely to disproportionately impact the measurement  of \ho  due to the currently small number of calibrator \sne.

\end{enumerate}

\section{Summary and Conclusions}
\label{sec:conclusions}

The infrared sensitivity and high resolution of \jwst offer a powerful new means of measuring the distances to nearby galaxies,  enabling independent determinations of \ho. In this study, we measured new distances to 10 nearby galaxies using \jwst observations of  TRGB and JAGB/carbon stars, all of which have well-observed \sne. The JAGB distance scale analysis, from  raw data to \ho determination, was conducted blindly. A comparison of the galaxy-to-galaxy 
TRGB and JAGB distances yielded  an average difference of less than 1\%. This level of consistency marks a significant advancement compared to recent decades.

A primary uncertainty in local \ho measurements using \sne ({\it e.g.,} \shoes or \cchp) stems from the limited number of nearby calibrating  \sn host galaxies. This scarcity arises from the rarity of \sne, and as a result, there are few galaxies in the local volume containing \sne that are also close enough to resolve Cepheids, TRGB or JAGB stars with \hst. Notably, many of these nearby supernovae are among the intrinsically brightest. To determine if a systematic error exists in distances to the more distant \sn galaxies, a larger sample is needed. Expanding this sample will mitigate potential bias and reduce statistical uncertainties. Unfortunately  these nearby \sne occur only every one to two years. Addressing the \ho tension will require a decrease in both the systematic and statistical uncertainties.

Our results, which are summarized in Figure \ref{fig:Ho_PDFs}, illustrate the agreement between our \cchp \hst + \jwst TRGB measurement applied to the \csp \sn sample and the cosmological \ho values from Planck \citep{planck_2018} and DESI \citep{DESI_2024}. The {\it accuracy} of our measurements is dominated by the systematic uncertainty in the distance to the anchor galaxy, \ngc 4258. Our current best estimate of \ho is  70.39 $\pm$ 1.22 (stat) $\pm$  1.33 (sys) $\pm$ 0.70 [$\sigma_{SN}$] \hounits. This value falls between that obtained from studies of the CMB, from BAO measurements calibrated by BBN, and the \shoes value, all without significant tension. While our results show consistency with $\Lambda$CDM, continued improvement to the local distance scale is essential for further reducing both systematic and statistical uncertainties.

\begin{acknowledgments}
This research is based in part on observations made with the NASA/ESA Hubble Space Telescope obtained from the Space Telescope Science Institute, which is operated by the Association of Universities for Research in Astronomy, Inc., under NASA contract NAS 5–26555. 
This work is also based in part on observations made with the NASA/ESA/CSA James Webb Space Telescope. 
The data were obtained from the Mikulski Archive for Space Telescopes at the Space Telescope Science Institute, which is operated by the Association of Universities for Research in Astronomy, Inc., under NASA contract NAS 5-03127.
These observations are associated with HST programs \#12880 and \#14149 and with JWST program \#1995.
Financial support for this work was provided in part by NASA through HST program \#16126 and JWST program \#1995.  AJL was supported by the Future Investigators in NASA Earth and Space Science and Technology (FINESST) award 80NSSC22K1602. 

This research has made use of the NASA/IPAC Extragalactic Database (NED), which is funded by the National Aeronautics and Space Administration and operated by the California Institute of Technology.

We thank Syed Uddin and Chris Burns for helpful advice on use of the CSP database and the MCMC analysis,  Mark Phillips for many years of collaboration as part of the CSP, and Eric Persson and Adam Riess for helpful conversations. We also thank Andy Dolphin and Dan Weisz and the \jwst Resolved Stellar Populations Early Release Science (ERS) team for developing the \nircam module of DOLPHOT, and for their advice on its implementation. In addition, WLF thanks Saul Perlmutter and Lloyd Knox for constructive comments on a draft version of this paper, as well as for helpful discussions. Many thanks to Marcia Rieke and the \nircam team, and to the many \jwst engineers for an extraordinary instrument and telescope.  Finally, we thank the University of Chicago and the Carnegie Institution for Science for their support of this research. 

The {\it JWST} imaging used in this paper can be found in MAST: \dataset[10.17909/ecf8-2z68]{http://dx.doi.org/10.17909/ecf8-2z68} and  http://dx.doi.org/10.17909/nz7r-mq84. The photometry of the JAGB stars for the galaxies in this program can be found at doi:\href{https://zenodo.org/records/14502265}{10.5281/zenodo.14502265}.

\end{acknowledgments}

%

\vspace{5mm}
\facilities{HST(WFC3,ACS,WFPC2), JWST(NIRCam) }


          
\software{Astropy \citep{astropy_2022}, DAOPHOT \citep{stetson_1987}, DOLPHOT \citep{dolphot_software}, {\it \bf emcee} \citep{foreman-mackey_2013}, Notebooks \citep{kluyver_jupyter_2016}, JWST Calibration Pipeline \citep{bushouse_2023}, Numpy  \citep{harris_numpy_2020}, Pandas \citep{mckinney_pandas_2010}, pymc \citep{pymc_2023}, SciPy \citep{virtanen_scipy_2020}}

\clearpage
\bibliography{Ho2024}{}
\bibliographystyle{aasjournal}
\appendix
\section{Three Additional Geometric Anchors for the JAGB Distance Scale} 
\label{App:AppendixA}

In the following we present a preliminary comparison of independent, geometry-based zero points for the JWST JAGB distance scale based on an interpolation between ground-based (2MASS) $J$-band (1.24 $\mu$m) data and space-based HST (F110W) data
for the same host-galaxy JAGB luminosity functions. 
From above and from below, these two filters closely bracket the \jwst/\nircam F115W filter, allowing a direct transformation from  the published ground-based JAGB calibrations onto the \jwst system.
The JAGB stars in both the Large and Small Magellanic Clouds are currently too bright for \jwst to image them without saturating. Below we offer a first look at an indirect solution to this problem: by defining a photometric transformation from both ground-based and \hst-based $J$-band photometry to the \jwst/\nircam system.

Figure A1 is a near-infrared (NIR) color-magnitude diagrams for the total population of high-luminosity resolved stars in M33 from the time-domain study of \citet{Javadi_2017}.
These are time-averaged data. 
Approximately 100 JAGB stars were identified and their marginalized luminosity function is shown in the right panel, where the mode of the JAGB population is found at J = 18.50 $\pm$ 0.006 (error on the mean). In a second, independent study of M33 \citep{lee_2022}  the mode of the (single-epoch) JAGB luminosity function for the outer-most region (shown in the right-most panel of their Figure 5) is found to be at J(2MASS) = 18.50 mag, with a quoted error on the mean of ±0.01 mag. The agreement between these two, totally independent datasets is reassuring.

A third NIR dataset for M33 comes from the HST study of the inner disk of this galaxy undertaken by \citet{Williams_2021} using WFC3-UV and -IR. In Figure A2 we show the NIR CMD for that dataset, where the J-band imaging was taken through their F110W filter, which is 0.14~$\mu$m to the blue of the 2MASS $J$-band filter that is centered at 1.24 $\mu$m. Our target wavelength, that of the F115W filter used by \jwst/\nircam is 1.15 $\mu$m. It is conveniently straddled by the two available datasets. 

\begin{figure}[ht!]
\figurenum{A1}
\plotone{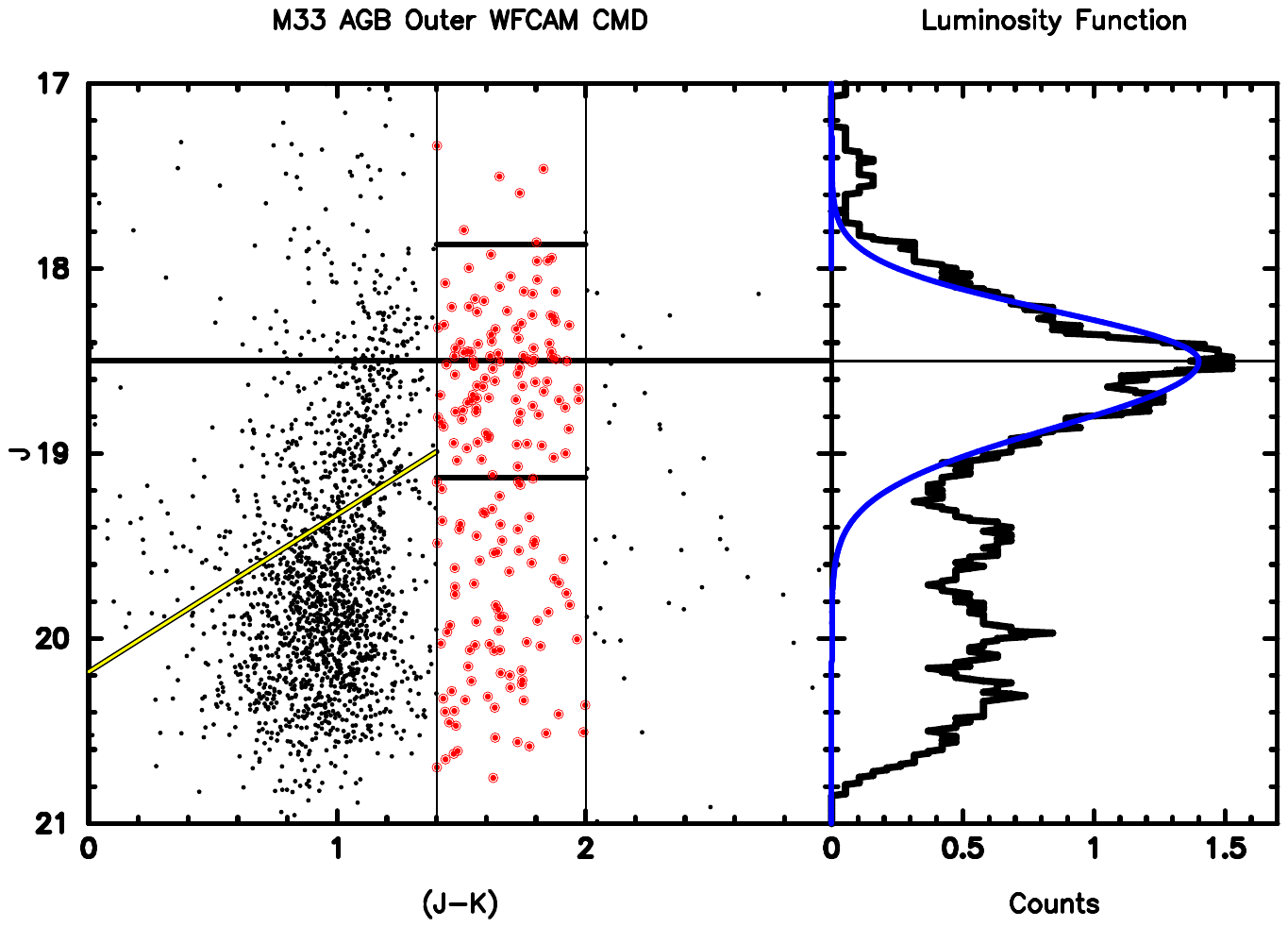}
\caption{Ground-Based NIR data for stars in the outer disk of M33: CMD (left panel) with color-selected JAGB stars shown in red, and the marginalized JAGB luminosity function (right panel) shown in black with an asymmetric gaussian fit shown in blue. Its modal value, at J = 18.50 mag is the blue line extrapolated back into the CMD to the left.}
\label{fig:M33-AGB-OUTER}
\end{figure}    

\begin{figure}[ht!]
\figurenum{A2}
\plotone{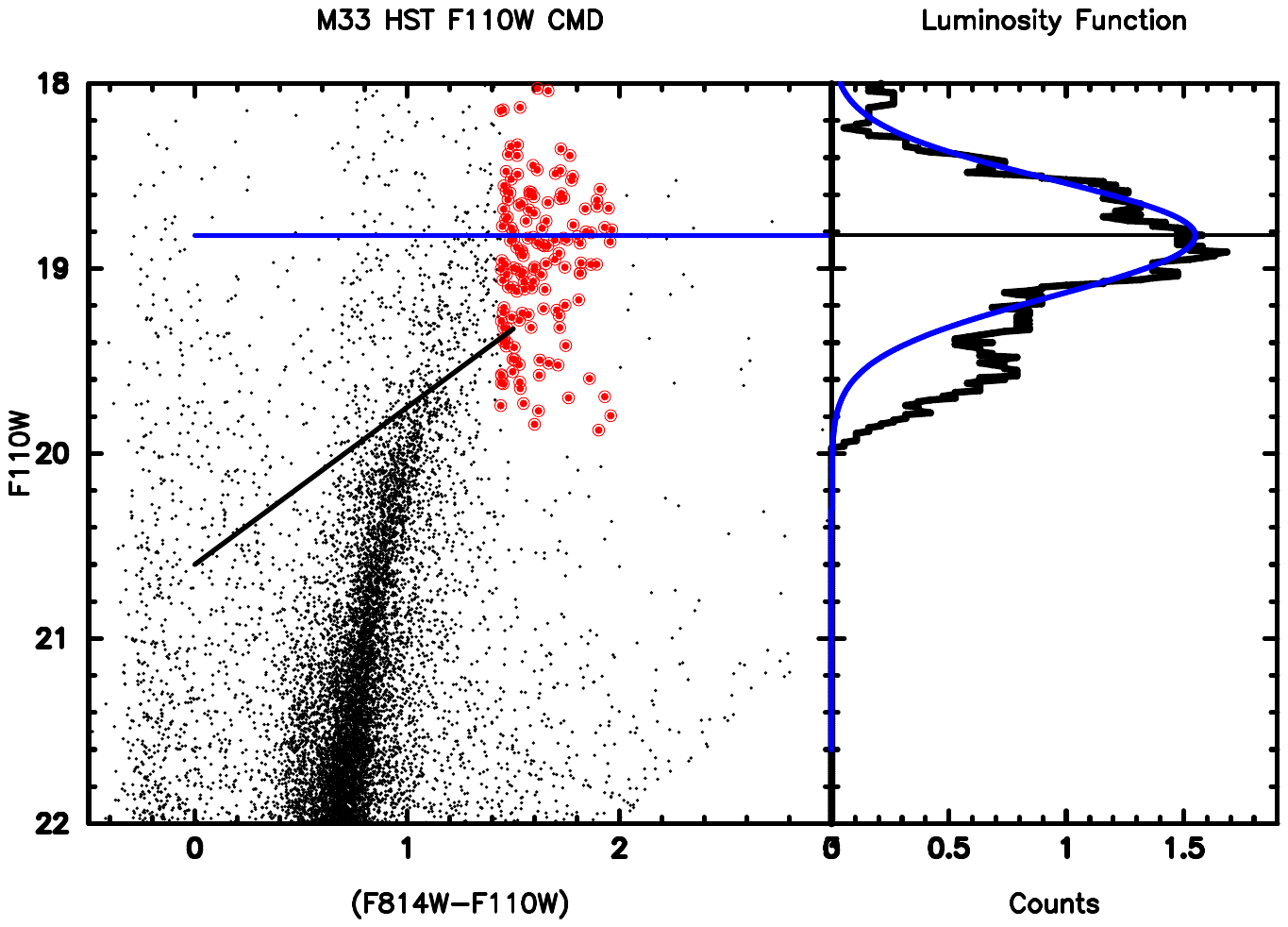}
\label{fig:M33-HST-CMD}
\caption{HST NIR data for stars in the outer disk of M33: CMD (left panel) with color-selected JAGB stars shown in red, and the marginalized JAGB luminosity function (right panel) shown in black with an asymmetric Gaussian fit shown in blue. Its modal value, at F110W = 18.82 mag is the blue line extrapolated back into the CMD to the left.}
\end{figure}


Figure A2 shows the HST CMD and JAGB luminosity function for M33. The mode of the JAGB population (color-selected, as shown in red) is found at J(F110W) = 18.82 $\pm$ 0.03 mag. (132 JAGB stars are found within the $\pm$0.64 (2-sigma) limits of the Gaussian fit, leading to an error on the mean of $\pm$0.027~mag.) The mode is fainter than the 2MASS value by +0.32 mag for the extended outer-disk population of JAGB stars in M33. A simple, linear interpolation between these two wavelengths gives a zero-point for a ground-based to JWST transformation tailored specifically to the AGB population, thereby avoiding the need to consider a color term. That transformation is J(F115W) = J(2MASS) + 0.205~mag.

\citet{madore_freedman_2020} give 2MASS absolute magnitude calibrations of the JAGB in the LMC \& SMC, as being M(2MASS) = -6.22 and -6.18 mag, respectively. These transform to M(JWST) -6.015 and -5.975~mag in the JWST/NIRCam F115W system. Additionally, we recall that a Milky Way (2MASS) absolute magnitude JAGB calibration, based on individual parallaxes, can be found in \citet{madore_2022}.   
Their averaged Milky Way value is $M(2MASS)$ = -6.19 $\pm$ 0.04 (stat), transforming to $M(JWST)$ = -5.985 $\pm$ 0.04~mag  

The global average of the three new (MW, LMC \& SMC) geometric anchors and including the NGC 4258 direct calibration of $M(JWST)$ = -5.99 $\pm$ 0.15~mag, gives $<M(JWST)>$ = -5.99 $\pm$0.015~mag (stat). This is -0.005~mag (marginally) brighter (0.25\% further in distance) than the value used in \citep{lee_2024a}, based on NGC~4258 alone, and presented again in the main text of this paper.

\section{Differences with Analyses in R24} 
\label{App:AppendixB}

Here we discuss a few points in response to those raised in R24.

\begin{enumerate}

\item{Method for correlated errors and H0:}

The analysis in the current paper differs substantially from an earlier preprint. We have postponed the measurement of a combined \ho value based on three methods until larger samples of JAGB and Cepheid distances are available.  The main results presented here are based on a significantly expanded TRGB galaxy sample. By combining our JWST TRGB results with earlier \hst measurements (F19, F20, F21) we have increased the sample from 10 galaxies to 20 galaxies, hosting 24 \sne. We have tabulated results for the smaller \jwst TRGB and JAGB samples individually, but note that  these values will improve as future \jwst observations become available.  To combine the \jwst TRGB and JAGB results, we have averaged the distances per host galaxy, rather than averaging final \ho values as  done previously. This approach ensures that covariances such as systematic errors from zero-point calibration are accurately propagated into the final \ho error budget.  

We note that our Cepheid analysis is still ongoing. In the disk fields where Cepheids reside, comparisons of different software packages (DAOPHOT and DOLPHOT) are showing significant photometric differences that require further investigation. These systematic differences are not present in the less crowded fields used for the TRGB analysis.

\item{ Method for measurement Uncertainties in NGC 4258 and Distance Comparisons:}

The  1.5\% geometric measurement uncertainty for \ngc 4258 was applied to each individual TRGB, JAGB and Cepheid measurement, and does not average down in a combined analysis. The 0.07  mag error for the Cepheids in our earlier work was a deliberate conservative estimate, reflecting ongoing uncertainties in the Cepheid measurements. While the R24 Cepheid distances are in accord with our TRGB distances,  the differences among Cepheid measurements (including our \jwst data and R24) are larger than those between the TRGB and JAGB distances. 

\item{Linearity of Cepheid Distances:}

As more distant galaxies are observed with \jwst, the linearity of the \hst distance scale will continue to be tested.  For example, there is a 3$\sigma$ correlation between M$_B$ and $\mu$  (shown in Figure \ref{fig:MBdistR22} below). Shown are the absolute peak B magnitudes, M$_B$, for the \shoes sample of 42 \sne as a function of distance modulus, all based on the Cepheid distance scale and \sn magnitudes from R22. All data points (blue and orange) are from the R22 data set. The orange points represent the subset of \sne observed as part of the current study. Our \jwst sample selection was based on proximity, to minimize potential systematics from crowding/blending effects.  

\medskip

A  3$\sigma$ trend is evident in Figure \ref{fig:MBdistR22}, in the sense that the supernova absolute magnitudes become fainter with increasing distance. The slope, measured using a Python orthogonal-regression algorithm, scipy.odr, is 0.082 $\pm$ 0.023, with a T-statistic  (the slope divided by its standard error) of 3.6.  This correlation is not an artifact of $\mu$  being present on both axes. For instance, a positive  +$\mu$ deviation would shift a point to the right (larger distance modulus) and upward (M = m – $\mu$ and M is negative), while a negative -$\mu$ deviation would shift it left and downward. Such deviations would not result in the observed slope. Future observations will determine whether this trend reflects a genuine distance-dependent effect. 

\medskip

We reiterate that four of the five galaxies compared in R24a, which concluded that crowding effects are ruled out at an 8.2-$\sigma$ confidence level, are at distances ~$\lesssim$ 20 Mpc. However, 60\% of the \shoes sample are at distances $>$ 20 Mpc. Therefore, future \jwst data are essential to complete this test. 

\end{enumerate}

\begin{figure*} 
\figurenum{B1}
 \centering
\includegraphics[width=1.0\textwidth]{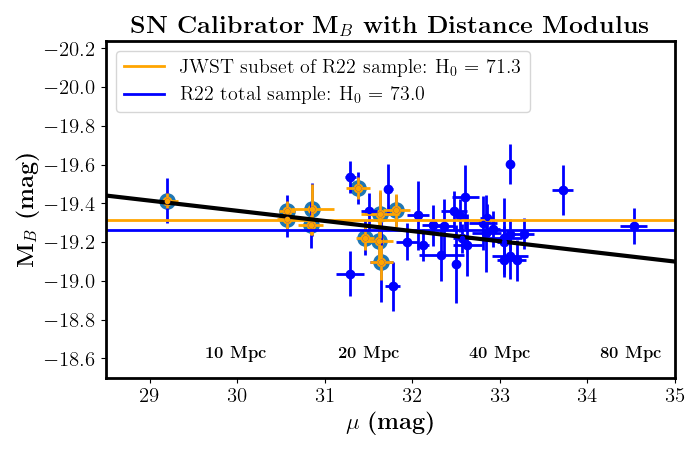}
 \caption{ Peak M$_B$ magnitudes versus distance modulus for \sne observed in galaxies with measured Cepheid distances from R22. The orange-filled blue circles  represent the subset of galaxies observed with \jwst as part of the current program. The black line is a weighted orthogonal distance regression fit with  3$\sigma$ significance. The orange and blue horizontal lines represent the weighted mean values of M$_B$ for the current \jwst and R22 total samples, respectively.  }
 \label{fig:MBdistR22}
\end{figure*}

\end{document}